\documentclass[aps,prb,preprint,showpacs,superscriptaddress,floatfix]{revtex4}
\usepackage{graphics}
\usepackage{epsfig}
\usepackage{amsmath}
\usepackage{amssymb}
\usepackage{calc}
\newcommand{\beq}{\begin{equation}}
\newcommand{\eeq}{\end{equation}}
\newcommand{\bea}{\begin{eqnarray}}
\newcommand{\eea}{\end{eqnarray}}
\DeclareMathOperator{\ee}{e}
\DeclareMathOperator{\dd}{d}
\DeclareMathOperator{\sgn}{sgn}

\begin{document}
\title{The attractor mechanism as a distillation procedure}
\author{P\'eter L\'evay}
\author{Szil\'ard Szalay}
\affiliation{Department of Theoretical Physics, Institute of Physics, Budapest University of Technology and Economics, H-1521 Budapest, Hungary}
\date{\today}
\begin{abstract}
In a recent paper it has been shown that for double extremal static spherical symmetric BPS black hole solutions in the STU model 
the well-known process of moduli stabilization at the horizon can be recast in a form of a distillation procedure of a three-qubit entangled state of GHZ-type.
By studying the full flow in moduli space in this paper we investigate this distillation procedure in more detail.
We introduce a three-qubit state with amplitudes depending on the conserved charges the warp factor, and the moduli. 
We show that for the recently discovered non-BPS solutions it is possible to see how the distillation procedure unfolds itself as we approach the horizon.
For the non-BPS seed solutions at the asymptotically Minkowski region we are starting with a three-qubit state having seven nonequal nonvanishing amplitudes and finally at the horizon we get a GHZ state 
with merely four nonvanishing ones with {\it equal} magnitudes. The magnitude of the surviving nonvanishing amplitudes is proportional to the macroscopic black hole entropy.
A systematic study of such attractor states shows that their properties reflect the structure of the fake superpotential. We also demonstrate that when starting with the very special values for the moduli corresponding to flat directions the uniform structure at the horizon deteriorates
due to errors generalizing the usual bit flips acting on the qubits of the attractor states.

\end{abstract}
\maketitle{}

\section{Introduction}

Recently striking multiple relations have been discovered between two seemingly unrelated fields: the physics of black hole solutions in string theory and the theory of quantum entanglement within quantum information 
theory\cite{Duff2,Linde,Levay1}.
Further papers established a complete dictionary between a variety of phenomena on one side of the correspondence in the language of the other.
This black hole qubit correspondence 
has repeatedly proved
to be useful for obtaining additional insight into both of the two fields\cite{DF1,Levay2,DF2,Levay3,Scherbakov,Borsten1,stu,Borsten2,Borsten3}.
The main correspondence found\cite{Duff2,Linde,DF1,Levay2,DF2,LVS} is between   the
macroscopic entropy
formulas obtained for certain black hole solutions in supergravity theories and
multiqubit and qutrit entanglement measures used in Quantum Information Theory.

Apart from understanding black hole entropy in quantum information theoretic terms the desire for an entanglement based understanding for issues of dynamics
also
arose.
In particular in the special case of the STU model\cite{Duff1,Behrndt,stu} it has been realized\cite{Levay1} that for extremal spherically symmetric BPS black hole solutions it is possible to rephrase the attractor mechanism\cite{attractor} as a distillation procedure of entangled "states" of very special kind on the event horizon.
Such states are of GHZ-type\cite{GHZ} or graph states\cite{graph,Levay3} well-known from quantum information theory.
The basic tool for establishing this result was the introduction of a three-qubit state $\vert\Psi\rangle$ depending on the conserved charges and also on the moduli fields.
Such "states" enjoy a number of remarkable properties\cite{Levay3}.
 The norm of this state having $8$ amlitudes is the black hole potential\cite{stu} $V_{BH}$. The flat covariant derivatives with respect to the K\"ahler connection are acting on $\vert\Psi\rangle$ as bit flip errors on the qubits.
At the horizon bit flip errors on $\vert\Psi\rangle$ are supressed for BPS solutions
and for non-BPS ones they are not. The non-BPS solutions can be characterized
by the number and types of bit-flip errors.

However, in these investigations\cite{Levay1,Levay3} establishing these results only double extremal\cite{Behrndt} solutions have been considered for which the moduli fields are constant even away from the horizon.
Since for this class of solutions $\vert\Psi\rangle$ is also constant
clearly within the context of such solutions it is not possible to get any additional insight on the important question how the distillation procedure unfolds itself as we approach the horizon after taking the limit $r\to 0$ with $r$ being the radial coordinate.

Luckily both in the BPS and non-BPS cases there exist more general static spherically symmetric solutions featuring the full radial flow in moduli space.
For the BPS case these are the well-known solutions based on harmonic functions\cite{Sabra} and for the non-BPS case
similar results generalizing these ones have recently become available\cite{Kubota,seed,Cai,stu}.
In these works after solving the equations of motion one obtains the attractor flow $z^j(r)$ in moduli space. Hence employing the charge and moduli dependent multiqubit states\cite{Levay1,Levay3} and using these solutions one might hope to get some additional insight into distillation issues by studying the corresponding flow $\vert\Psi(r)\rangle$.

The aim of the present paper is to investigate this distillation procedure in detail for  the special case of extremal spherically symmetric black hole solutions in the STU model.
We will use a special combination of the moduli fields, the warp factor and the conserved charges reminiscent of a $3$-qubit state of Quantum Information Theory. We will call this creature a "three-qubit state" furnishing a representation space for the action of the duality group $\mathrm{SL}(2,\mathbb R)^{\times 3}\subset \mathrm{Sp}(8,\mathbb R)$, though this terminology might be misleading.
It will be obvious that our state has intimate connections to entities like the "fake superpotential"\cite{fake} and even possibly to the phase of the semiclassical wave function used in recent studies\cite{Pioline}, however in this paper we will not elaborate on its physical meaning.
Results on the origin of these $3$-qubit states having some relevance on such interesting issues will be presented in an acompanying paper\cite{Levay4}.

The organization of this paper is as follows.
In Section II. we summarize the usual formalism of the STU model.
In Section III. we introduce our moduli and charge-dependent $3$-qubit state, and recall results concerning the black hole qubit correspondence that we will need later. In Section IV. we reformulate the well-known findings concerning BPS solutions based on harmonic functions. Here we show that the "attractor at infinity"\cite{infti,stu} corresponds to a distillation procedure of a {\it normalized} GHZ state (dual to the usual one at the horizon\cite{Levay1}) at the asymptotically Minkowski region.
In Section V. we study the flow $\vert\Psi(r)\rangle$ for the non-BPS $D0-D4$ system answering the seed solution\cite{seed}.
Here we generalize further our $3$-qubit state by including also the warp factor into its definition.
We show that the Fourier amplitudes of this state in the discrete Fourier (Hadamard) transformed
basis satisfy a set of first order differential equations. 
Using the results of the previous sections in Section VI. for the non-BPS seed solutions we demonstrate 
how a GHZ state at the horizon emerges from a state characterizing the flow at the asymptotically Minkowski region.   
The attractor mechanism in this picture simply amounts to the fact that three amplitudes out of the seven nonequal nonvanishing ones are dying out as we approach the horizon.
The remaining amplitudes have the same magnitudes related to the macroscopic black hole entropy. The relative phase factors of these amplitudes are merely signs reflecting the structure of the fake superpotential\cite{fake,stu}.
In Section VII. using recent general results on the STU model\cite{stu} we calculate the explicit form of the states at the horizon for both the available BPS and non-BPS solutions.
Featuring the parametrization\cite{seed,stu} revealing the flat directions\cite{flat} 
we show that the role of the flat directions in this picture is to deteriorate the uniform GHZ-like structure on the horizon.
According to the results of our previous paper\cite{Levay3}
the differences between different types of solutions (BPS, non-BPS with vanishing \cite{Scherbakov} and nonvanishing\cite{Soroush} central charge) manifest themselves in applying bit flip errors to the relevant GHZ-like state of the BPS flow. 
In view of this result flat directions give a new twist on this picture, namely when starting the flow in one of the flat directions the resulting state on the horizon will exhibit errors of more general type than the usual bit flip ones.
Finally in Section VIII. we present our conclusions, with some calculational details left to an Appendix.

\section{STU black holes}

In the following we consider ungauged $N=2$ supergravity in $d=4$ coupled to
$n$ vector multiplets.
The $n=3$ case corresponds to the $STU$ model.
The bosonic part of the action (without hypermultiplets) (in units of $G_N=1$) is

\begin{eqnarray}
\label{Action}
{\cal S}&=&\frac{1}{16\pi}\int \dd^4x\sqrt{\vert g\vert }\{-\frac{R}{2}+G_{a\overline{b}}{\partial}_{\mu}z^a{\partial}_{\nu}{\overline{z}}^{\overline{b}}g^{\mu\nu}\nonumber\\&+&({\rm Im}{\cal N}_{IJ}{\cal F}^I\cdot{\cal F}^J+{\rm Re}{\cal N}_{IJ}{\cal F}^I\cdot{^\ast{\cal F}^J})\}.
\end{eqnarray}
Here ${\cal F}^I$, and ${^\ast{\cal F}^I}$,  $I=0,1,2\dots n$ are two-forms associated to the field strengths ${\cal F}^I_{\mu\nu}$  of $n+1$ $\mathrm{U}(1)$ gauge-fields and their duals.
The $z^a$ $a=1,\dots n$ are complex scalar (moduli) fields that can be regarded as local coordinates on a projective special K\"ahler manifold ${\cal M}$. This manifold for the STU model is
$\mathrm{SL}(2, \mathbb R)/\mathrm{U}(1)^{\times 3}$.
In the following we will denote the three complex scalar fields as
\beq
z^a\equiv x^a-iy^a,\qquad y^a>0,\qquad a=1,2,3.
\label{moduli}
\eeq
\noindent
With these definitions the metric and the connection on the scalar manifold are
\beq
G_{a\overline{b}}=\frac{\delta_{a\overline{b}}}{(2y^a)^2},\qquad
\qquad{\Gamma}^{a}_{aa}=\frac{-i}{y^a}.
\label{targetmetric}
\eeq
\noindent
The metric above can be derived from the K\"ahler potential
\beq
K= -\log(8y^1y^2y^3)
\label{Kahler}
\eeq
\noindent
as $G_{a\overline{b}}={\partial}_a{\partial}_{\overline{b}}K$.
For the STU model the scalar dependent vector couplings ${\rm Re}{\cal N}_{IJ}$
and ${\rm Im}{\cal N}_{IJ}$ take the following form
\beq
{\rm Re}{\cal N}_{IJ}=\begin{pmatrix}2x^1x^2x^3&-x^2x^3&-x^1x^3&-x^1x^2\\-x^2x^3&0&x^3&x^2\\-x^1x^3&x^3&0&x^1\\-x^1x^2&x^2&x^1&0\end{pmatrix},
\label{valos}
\eeq 
\beq
{\rm Im}{\cal N}_{IJ}=-y^1y^2y^3\begin{pmatrix}1+{\left(\frac{x^1}{y^1}\right)}^2+{\left(\frac{x^2}{y^2}\right)}^2+{\left(\frac{x^3}{y^3}\right)}^2&-\frac{x^1}{(y^1)^2}&-\frac{x^2}{(y^2)^2}&-\frac{x^3}{(y^3)^2}\\-\frac{x^1}{(y^1)^2}&\frac{1}{(y^1)^2}&0&0\\-\frac{x^2}{(y^2)^2}&0&\frac{1}{(y^2)^2}&0\\-\frac{x^3}{(y^3)^2}&0&0&\frac{1}{(y^3)^2}\end{pmatrix}.
\label{kepzetes}
\eeq
\noindent
We note that these vector couplings can be derived from the
holomorphic prepotential
\beq
F(X)=\frac{X^1X^2X^3}{X^0},\qquad X^I=(X^0,X^0z^a),
\eeq
\noindent
via the standard procedure characterizing special K\"ahler geometry\cite{Strominger}.

For the physical motivation of Eq.~(\ref{Action}) we note that when type IIA string theory is compactified on a $T^6$ of the form $T^2\times T^2\times T^2$ one recovers $N=8$ supergravity in $d=4$ with $28$ vectors and $70$ scalars taking values in the symmetric space $\mathrm{E}_{7(7)}/\mathrm{SU}(8)$. This $N=8$ model with an on shell U-duality symmetry $\mathrm{E}_{7(7)}$ has a consistent $N=2$ truncation with $4$ vectors and three complex scalars which is just the STU model\cite{Duff1,Behrndt}. 
The $D0-D2-D4-D6$ branes wrapping the various $T^2$ give rise to four electric and four magnetic charges defined as
\beq
P^I=\frac{1}{4\pi}\int_{S^2}{\cal F}^I,\qquad
Q_I=\frac{1}{4\pi}\int_{S^2}{\cal G}_I,\quad I=0,1,2,3
\label{charges}
\eeq
\noindent
where
\beq
{\cal G}_I=\overline{\cal N}_{IJ}{\cal F}^{+I},\qquad {\cal F}^{\pm I}_{\mu\nu}={\cal F}^I_{\mu\nu}\pm \frac{i}{2}{\varepsilon}_{\mu\nu\rho\sigma}{\cal F}^{I\rho\sigma}.
\label{G}
\eeq
\noindent
These charges can be organized into symplectic pairs 
\beq
{\Gamma}\equiv(P^I, Q_J)
\label{Gamma}
\eeq
\noindent
and have units of length. They are related to the dimensionless quantized charges by some dressing factors. Using the variable $\tau=1/r=1/\vert{\bf x}\vert$ and normalizing the asymptotic moduli as
\beq
y^a(0)=1\qquad x^a(0)=B^a
\label{normmagnetic}
\eeq
\noindent
 the dressing factors are essentially the masses of the underlying branes\cite{seed}.

We are interested in static, spherically symmetric, extremal black hole solutions associated to the (\ref{Action}) action.
The ansatz for the metric is
\beq
\dd s^2=-e^{2U(\tau)}\dd t^2+e^{-2U(\tau)}\dd{\bf x}^2
\label{ansatz}
\eeq
\noindent
where the warp factor is a function of $\tau=1/r$. 
Putting this ansatz into (\ref{Action}) we obtain a one dimensional effective Lagrangian for the radial evolution of the quantities $U(\tau)$, $z^a(\tau)$, as well as the electric and magnetic potentials\cite{Kallosh2}
\beq
{\cal L}(U(\tau), z^a(\tau),\overline{z}^{\overline{a}}(\tau))=\left(\frac{\dd U}  {\dd\tau}\right)^2+G_{a\overline{a}}\frac{\dd z^a}{\dd\tau}\frac{\dd\overline{z}^{\overline{a}}}{\dd\tau}+e^{2U}V_{BH}(z,\overline{z},P,Q),
\label{Lagrange}
\eeq
\noindent
and the constraint
\beq
\left(
\frac{\dd{\cal U}}{\dd\tau}\right)^2+G_{a\overline{a}}\frac{\dd z^a}{\dd\tau}
\frac{\dd\overline{z}^{\overline{a}}}{\dd\tau}-e^{2U}V_{BH}(z,\overline{z},P,Q)=0.
\label{constraint}
\eeq
Here our quantity of central importance is the black hole potential $V_{BH}$
which is depending on the moduli as well on the charges. 
Its explicit form is given by
\beq
V_{BH}=\frac{1}{2}\begin{pmatrix}P^I&Q_I\end{pmatrix}\begin{pmatrix}(\mu+\nu{\mu}^{-1}\nu)_{IJ}&-(\nu{\mu}^{-1})^J_I\\-({\mu}^{-1}\nu)^I_J&({\mu}^{-1})^{IJ}\end{pmatrix}\begin{pmatrix}P^J\\Q_J\end{pmatrix},
\label{potential}
\eeq
\noindent
where the matrices ${\nu}={\rm Re}{\cal N}$ and ${\mu}={\rm Im}{\cal N}$ are the ones of Eqs.~(\ref{valos}) and (\ref{kepzetes}).
The explicit form of ${\mu}^{-1}$ is 
\beq
{\mu}^{-1}=\frac{-1}{y^1y^2y^3}\begin{pmatrix}1&x^1&x^2&x^3\\x^1&{\vert z^1\vert}^2&x^1x^2&x^1x^3\\x^2&x^1x^2&{\vert z^2\vert}^2&x^2x^3\\x^3&x^1x^3&x^2x^3&{\vert z^3\vert}^2\end{pmatrix}.
\eeq
An alternative expression for $V_{BH}$ can be given in terms of the central charge of $N=2$ supergravity, i.e.~the charge of the graviphoton.
\beq
V_{BH}=Z\overline{Z}+G^{a\overline{b}}(D_aZ)({\overline{D}}_{\overline{b}}\overline{Z})
\eeq
\noindent
where for the STU model
\beq
Z=e^{K/2}W=e^{K/2}(Q_0+z^1Q_1+z^2Q_2+z^3Q_3+z^1z^2z^3P^0-z^2z^3P^1-z^1z^3P^2-z^1z^2P^3),
\label{central}
\eeq
and $D_a$ is the K\"ahler covariant derivative 
\beq
D_aZ=({\partial}_a+\frac{1}{2}{\partial}_aK)Z,
\label{kovder}
\eeq
\noindent
and $W$ is the superpotential.

Extremization of the effective Lagrangian Eq.~(\ref{Lagrange}) with respect to the warp factor and the scalar fields yields the Euler-Lagrange equations
\beq 
\ddot{U}=e^{2U}V_{BH},\qquad \ddot{z}^{a}+\Gamma^a_{bc}\dot{z}^b\dot{z}^c=e^{2U}{\partial}^aV_{BH}.
\label{Euler}
\eeq
\noindent
In these equations the dots denote derivatives with respect to $\tau$.
These equations taken together with the constraint Eq.~(\ref{constraint})
determine the black hole solutions whose quantum information theoretic interpretation we are interested in.

\section{Three-qubit states}

It is useful to reorganize the charges of the STU model into the $8$ amplitudes of a three-qubit state
\beq
\vert\Gamma\rangle=\sum_{l,k,j=0,1}{\Gamma}_{lkj}\vert lkj\rangle\qquad \vert lkj\rangle\equiv\vert l\rangle_3\otimes\vert k\rangle_2\otimes\vert j\rangle_1
\label{chargegamma}
\eeq
\noindent
where
\beq
\frac{1}{\sqrt{2}}\begin{pmatrix}P^0,&P^1,&P^2,&P^3\\-Q_0,&Q_1,&Q_2,&Q_3\end{pmatrix}=
\begin{pmatrix}{\Gamma}_{000},&{\Gamma}_{001},&{\Gamma}_{010},&{\Gamma}_{100}\\{\Gamma}_{111},&{\Gamma}_{110},&{\Gamma}_{101},&{\Gamma}_{011}\end{pmatrix}. 
\label{cgamma}
\eeq
\noindent
Notice that we have introduced the convention of labelling the qubits from the right to the left.   
Moreover, for convenience we have also included a factor $\frac{1}{\sqrt{2}}$ into our definition. The state $\vert\Gamma\rangle$ is a three-qubit state of a very special kind.
First of all this state defined by the charges need not have to be normalized. Moreover, the amplitudes of this state are not complex numbers but {\it real} ones.
As a next step we can define a new three-qubit state $\vert\Psi\rangle$ depending on the charges $\Gamma$
and
also on the moduli\cite{Levay1,Levay3}. This new state will be a three-qubit state with $8$ complex amplitudes. However, as we will see it is really a {\it real three-qubit state},
since it is $\mathrm{U}(2)^{\times 3}$ equivalent to a one with $8$ real amplitudes. 

In order to motivate our definition of the new state $\vert\Psi\rangle$ we notice that\cite{Levay3}
\beq
V_{BH}=\frac{1}{y^1y^2y^3}\langle\Gamma\vert\begin{pmatrix}{\vert z^3\vert}^2&-z^3\\-z^3&1\end{pmatrix}\otimes\begin{pmatrix}{\vert z^2\vert}^2&-z^2\\-z^2&1\end{pmatrix}\otimes\begin{pmatrix}{\vert z^1\vert}^2&-z^1\\-z^1&1\end{pmatrix}\vert\Gamma\rangle.
\eeq
\noindent
Now we define the state $\vert\Psi\rangle$ as
\beq
\vert\Psi(z^a,\overline{z}^{\overline{a}},\Gamma)\rangle =
e^{K/2}\begin{pmatrix}\overline{z}^3&-1\\-z^3&1\end{pmatrix}\otimes
\begin{pmatrix}\overline{z}^2&-1\\-z^2&1\end{pmatrix}\otimes\begin{pmatrix}\overline{z}^1&-1\\-z^1&1\end{pmatrix}\vert\Gamma\rangle.
\label{Psi}
\eeq
\noindent
Introducing the matrices
\beq
{\cal S}_a\equiv\frac{1}{\sqrt{2y^a}}\begin{pmatrix}\overline{z}^a&-1\\-z^a&1\end{pmatrix}={\cal U}S_a\equiv\frac{1}{\sqrt{2}}\begin{pmatrix}i&-1\\i&1\end{pmatrix}\frac{1}{\sqrt{y^a}}\begin{pmatrix}y^a&0\\-x^a&1\end{pmatrix},\qquad a=1,2,3.
\label{Smatrix}
\eeq
\noindent
With this notation we have 
\beq
\vert\Psi\rangle=({\cal S}_3\otimes {\cal S}_2\otimes {\cal S}_1) \vert\Gamma\rangle=({\cal U}\otimes {\cal U}\otimes {\cal U})(S_3\otimes S_2\otimes S_1)\vert\Gamma\rangle.
\label{3qubitallapot}
\eeq
This means that the state $\vert\Psi\rangle$ up to a phase for all values of the moduli
is in the $\mathrm{GL}(2, {\bf C})^{\times 3}$ orbit of the charge state $\vert\Gamma\rangle$. 
Obviously the state $\vert\Psi\rangle$ is an unnormalized three-qubit one
with $8$ complex amplitudes. However, it is {\it not} a genuine complex three-qubit state but rather a one which is $\mathrm{U}(2)^{\times 3}$ equivalent to a real one. This should not come as a surprise 
since the symmetry group associated with the STU model is not $\mathrm{GL}(2, \mathbb C)^{\times 3}$ but rather $\mathrm{SL}(2,\mathbb R)^{\times 3}$.

Using these definitions we can write the black hole potential in the following nice form
\beq
V_{BH}={\vert\vert\Psi\vert\vert}^2.
\label{Norma}
\eeq
\noindent
Here the norm is defined using the usual scalar product in ${\mathbb C}^8\simeq {\mathbb C}^2\otimes {\mathbb C}^2\otimes{\mathbb C}^2$ with complex conjugation in the first factor. Since the norm is invariant under $\mathrm{U}(2)^{\times 3}$ our choice of the first unitary matrix of Eq.~(\ref{Smatrix}) is not relevant in the structure of $V_{BH}$.
We could have defined a new moduli dependent real state instead of the complex one $\vert\Psi\rangle$ by using merely the $\mathrm{SL}(2, {\mathbb R})$ matrices
of Eq.~(\ref{Smatrix}) for their definition.
However, we prefer the complex form of Eq.~(\ref{3qubitallapot}) since it will be useful later.

It is instructive to write out explicitly the amplitudes of our complex three-qubit state $\vert\Psi\rangle$.
\beq
\sqrt{2}{\Psi}_{000}=e^{K/2}W(\overline{z}^3,\overline{z}^2,\overline{z}^1),\qquad \sqrt{2}{\Psi}_{111}=-e^{K/2}W(z^3,z^2,z^1),
\label{000}
\eeq
\beq
\sqrt{2}{\Psi}_{110}=e^{K/2}W(z^3,z^2,\overline{z}^1),\qquad
\sqrt{2}{\Psi}_{001}=-e^{K/2}W(\overline{z}^3,\overline{z}^2,z^1),
\label{001}
\eeq
\noindent
with the remaining amplitudes arising by cyclic permutation.
Notice also that we have the property (reality condition)
\beq
{\Psi}_{000}=-\overline{\Psi}_{111},\quad 
{\Psi}_{110}=-\overline{\Psi}_{001},\quad
{\Psi}_{101}=-\overline{\Psi}_{010},\quad
{\Psi}_{011}=-\overline{\Psi}_{100},
\label{conj}
\eeq
\noindent 
which can be written as
\beq
\vert\overline{\Psi}\rangle+(\sigma_1\otimes \sigma_1\otimes \sigma_1)\vert\Psi\rangle =0
\label{reality}
\eeq
\noindent
via the bit flip operator ${\sigma}_1$.
Using this in Eq.~(\ref{Norma}) we can write $V_{BH}$ in the alternative form\cite{Soroush}
\beq
V_{BH}=e^K\left({\vert W(z^3,z^2,z^1)\vert}^2+{\vert W(z^3,z^2,\overline{z}^1)\vert}^2+
{\vert W(z^3,\overline{z}^2,z^1)\vert}^2+{\vert W(\overline{z}^3,z^2,z^1)\vert}^2\right).
\eeq
\noindent

As a motivation for our particular definition for $\vert\Psi\rangle$ we
notice that 
\beq
\sqrt{2}\Psi_{000}=\overline{Z},
\label{centralcharge}
\eeq
i.e.~the amplitudes ${\Psi}_{111}$ and $\Psi_{000}$ are related to the central charge and its complex conjugate.
The remaining amplitudes are simply arising by conjugating one or two moduli.
Moreover, one can show\cite{Levay3} that the flat covariant derivatives
with respect to the moduli act on our $\vert\Psi\rangle$ as bit flip errors.
Explicitly we have
\beq
{D}_{\hat{1}}\vert\Psi\rangle=(I\otimes I\otimes \sigma_{+})\vert\Psi\rangle
,\qquad {D}_{\hat{\overline{1}}}\vert\Psi\rangle=
(I\otimes I\otimes \sigma_{-})\vert\Psi\rangle,\nonumber
\eeq
\beq
D_{\hat{2}}\vert\Psi\rangle=(I\otimes \sigma_{+}\otimes I)\vert\Psi\rangle
,\qquad {D}_{\hat{\overline{2}}}\vert\Psi\rangle=
(I\otimes \sigma_{-}\otimes I)\vert\Psi\rangle,
\label{flatkovariantder}
\eeq
\beq
{D}_{\hat{3}}\vert\Psi\rangle=(\sigma_{+}\otimes I\otimes I)\vert\Psi\rangle
,\qquad {D}_{\hat{\overline{3}}}\vert\Psi\rangle=
(\sigma_{-}\otimes I\otimes I)\vert\Psi\rangle.\nonumber
\eeq
\noindent
Here
the operators $\sigma_{\pm}$ act as
\beq
\sigma_{+}\vert 0\rangle=\vert1\rangle,\qquad \sigma_{+}\vert 1\rangle =0,  \qquad \sigma_{-}\vert 0\rangle=0,\qquad \sigma_{-}\vert 1\rangle=\vert 0\rangle,
\eeq 
\noindent
and
the flat covariant derivatives are 
defined as $D_{\hat{1}}=-2iy^1D_1$, 
$D_{\hat{\overline{1}}}=2iy^1D_{\overline{1}}$
e.t.c. where
\beq
D_1W(z^3,z^2,z^1)=\frac{W(z^3,z^2,\overline{z}^1)}{\overline{z}^1-z^1},\qquad
D_{\overline{1}}W(z^3,z^2,z^1)=0\qquad{\rm e.t.c.}
\eeq
\noindent
Hence the flat covariant derivatives are acting on our three-qubit state $\vert\Psi\rangle$ as the operators of {\it projective errors} known from the theory of quantum error correction.
Alternatively one can look at the action of the combination
$D_{\hat{a}}+D_{\hat{\overline{a}}}$
\beq
(D_{\hat{1}}+D_{\hat{\overline{1}}})\vert\Psi\rangle =(I\otimes I\otimes \sigma_1)\vert\Psi\rangle,\quad{\rm e.t.c.}
\label{flipcov}
\eeq
\noindent

According to our previous paper\cite{Levay3} it is illuminating to
use the discrete Fourier transform of our three-qubit state $\vert\Psi\rangle$.
At the event horizon the moduli are stabilized due to the attractor mechanism, and their stabilized values can be expressed in terms of the charges.
These stabilized values give rise to entangled states on the horizon of very special form. It was shown\cite{Levay1,Levay3} that for BPS solutions these states or of generalized Greenberger-Horne-Zeilinger (GHZ) form\cite{GHZ}, and for the simple non-BPS solutions\cite{Soroush} they are graph-states\cite{graph} well-known from quantum information theory.

The discrete Fourier (Hadamard) transformation is implemented
by acting on $\vert\Psi\rangle$ by $H\otimes H\otimes H$ where
\beq
H=\frac{1}{\sqrt{2}}\begin{pmatrix}1&1\\1&-1\end{pmatrix}.
\label{Hadamard}
\eeq
\noindent
Hence the Fourier transformed basis states are defined as
\beq
\label{HadamardBase}
\vert \tilde{0}\rangle\equiv\frac{1}{\sqrt{2}}(\vert 0\rangle +\vert 1\rangle)=H\vert 0\rangle,\qquad \vert\tilde{1}\rangle\equiv\frac{1}{\sqrt{2}}(\vert 0\rangle -\vert 1\rangle) =H
\vert 1\rangle.
\eeq
\noindent
Since $H\sigma_1H=\sigma_3$ and $H\sigma_3H=\sigma_1$ the operator $\sigma_1$ is acting on the Hadamard
transformed base as a phase (sign) flip operator and vice versa.
The important corollary of this observation is that in the theory of quantum    error
correction once we have found a means for correcting bit flip errors using a
discrete Fourier transform the same technique can be used for correcting phase  flip ones.
It was shown\cite{Levay3} that the phase flip errors in the discrete Fourier transformed base
correspond to flipping the sign of certain charges of the symplectic vector $\Gamma$.
For BPS solutions these phase flips are supressed, and for the simple non-BPS solutions only errors of very special kind are allowed.

Now we would like to gain some more insight into these interesting results by studying the solutions even away from the horizon. For this the full solution
of the flow in moduli space is needed. For BPS solutions we can use the well-known 
results\cite{Sabra,Denef} and for the non-BPS solutions the recently found seed solutions\cite{Kubota,seed,Cai} and the most general non-BPS flows\cite{stu} generalizing the simple non-BPS solutions\cite{Cvetic,Soroush,Horowitz}. 

Our main calculational tool will be to consider the discrete Fourier transform of our $\vert\Psi\rangle$ 
\beq
\vert\tilde{\Psi}\rangle=(H\otimes H\otimes H)\vert\Psi\rangle=({\cal P}\otimes{\cal P}\otimes {\cal P})(S_3\otimes S_2\otimes S_1)\vert\Gamma\rangle
\label{Fourier}
\eeq
\noindent
where
\beq
{\cal P}=\begin{pmatrix}i&0\\0&-1\end{pmatrix}
\label{phasegate}
\eeq
\noindent
is just $i$ times the usual phase gate known from quantum information theory.
Hence from Eq.~(\ref{Fourier}) we see that the Fourier transformed state is up to some important complex phase factors is lying on the $\mathrm{SL}(2,{\mathbb R})^{\times 3}$ orbit of the charge state $\vert\Gamma\rangle$.
Now solving Eqs.~(\ref{Euler}) with the constraint of Eq.~(\ref{constraint}) yields the flow $z^a(\tau)$ on moduli space. To this flow we can associate a corresponding one $\vert\tilde{\Psi}(\tau)\rangle$ of the Fourier transformed state.
Our aim for the following sections is to look at the structure of $\vert\tilde{\Psi}(\tau)\rangle$
for BPS and the non-BPS seed solutions.

\section{BPS solutions}

In this section we would like to study the behavior of BPS solutions in the three-qubit picture.
In particular we would like to see how the three-qubit state of Eq.~(\ref{Psi})
behaves as a function of ${\tau}$, answering to the flow $z^a(\tau)$ in moduli 
space.
This three-qubit picture is natural as the $U$-duality group is $\mathrm{SL}(2,\mathbb R)^{\times 3}$ which is a subgroup of $\mathrm{Sp}(8,\mathbb R)$, hence we expect that the usual symplectic invariants occurring in the formalism of the STU model should boil down to the corresponding $\mathrm{SL}(2,\mathbb R)^{\times 3}$ i.e.~three-qubit ones.

As a first step in order to present the solutions capable of incorporating a wider range of asymptotic data with $B$-fields we define a set of harmonic functions as
\beq
{\cal H}(\tau)=\overline{\Gamma}+{\Gamma}\tau,\qquad {\rm i.e.}\qquad H^I=\overline{P}^I+P^I\tau,\quad H_J=\overline{Q}_J+Q_J\tau\quad I,J=0,1,2,3.
\label{Harmonic}
\eeq
\noindent
We can alternatively encode the asymptotic data (${\cal H}(0)=\overline{\Gamma}$)
into a three-qubit state 
\beq
\vert\overline{\Gamma}\rangle=\sum_{l,k,j=0,1}\overline{\Gamma}_{lkj}\vert lkj\rangle,
\qquad
\begin{pmatrix}\overline{\Gamma}_{000},&\overline{\Gamma}_{001},&\overline{\Gamma}_{010},&\overline{\Gamma}_{100}\\\overline{
\Gamma}_{111},&\overline{\Gamma}_{110},&\overline{\Gamma}_{101},&\overline{\Gamma}_{011}\end{pmatrix}
\equiv
\frac{1}{\sqrt{2}}\begin{pmatrix}\overline{P}^0,&\overline{P}^1,&\overline{P}^2,&\overline{P}^3\\-\overline{Q}_0,&\overline{Q}_1,&\overline{Q}_2,&\overline{Q}_3\end{pmatrix}.
\label{overgamma}
\eeq
\noindent
which plays a role similar to the charge state of Eq.~(\ref{chargegamma}).
Notice that unlike for $\overline{\Gamma}$ in Eq.~(\ref{Harmonic}) in the state $\vert\overline{\Gamma}\rangle$ we also included a factor of $\frac{1}{\sqrt{2}}$ for convenience.

For later use we also define a $\tau$ dependent three-qubit state
as
\beq
\vert{\cal H}(\tau)\rangle \equiv \vert\overline{\Gamma}\rangle+\tau\vert\Gamma\rangle.
\label{harmonicstate}
\eeq
\noindent
This state in the limits $\tau\rightarrow 0$ and $\tau\rightarrow\infty$
characterizes the asymptotic data and the charge configuration respectively.

Let us now define Cayley's hyperdeterminant\cite{Cayley,Kundu} $D(\vert\psi\rangle)$ for an arbitrary three-qubit state
$\vert \psi\rangle=\sum_{lkj=0,1}{\psi}_{lkj}\vert lkj\rangle$ with amplitudes
\beq
\begin{pmatrix}{\psi}_{000},&{\psi}_{001},&{\psi}_{010}
,&{\psi}_{100}\\
{\psi}_{111},&{\psi}_{110},&{\psi}_{101},&{\psi
}_{011}\end{pmatrix}\equiv
\begin{pmatrix}{\psi}_0,&{\psi}_1,&{\psi}_2,&{\psi}_4\\
{\psi}_7,&{\psi}_6,&{\psi}_5,&{\psi}_3\end{pmatrix}
\eeq
\noindent
as
\begin{eqnarray}
D(\vert\psi\rangle)\equiv(\psi_0\psi_7)^2&+&(\psi_1\psi_6)^2+(\psi_2\psi_5)^2+(\psi_4\psi_3)^2
-2(\psi_0\psi_7)[(\psi_1\psi_6)+(\psi_2\psi_5)+(\psi_4\psi_3)]\nonumber\\-
2[(\psi_1\psi_6)(\psi_2\psi_5)&+&(\psi_1\psi_6)(\psi_4\psi_3)+(\psi_2\psi_5)(\psi_4\psi_3)]+4\psi_0\psi_1\psi_2\psi_4+4\psi_7\psi_6\psi_5\psi_3.
\label{Cayley}
\end{eqnarray}
\noindent
$D(\vert\psi\rangle)$ is permutation and $\mathrm{SL}(2,{\mathbb C})^{\times 3}$ invariant and under the group $\mathrm{GL}(2, {\mathbb C})^{\times 3}$ transforms as
\beq
D(\psi)\mapsto ({\rm Det}G_3)^2({\rm Det}G_2)^2({\rm Det}G_1)^2 D(\psi),\qquad
G_3\otimes G_2\otimes G_1\in \mathrm{GL}(2, {\mathbb C})^{\times 3}.
\label{relativeinv}
\eeq
Notice that for a general charge state $\vert\Gamma\rangle$ such as the one of Eq.~(\ref{chargegamma}) we have
\beq
-4D(\vert\Gamma\rangle)=I_4(\Gamma)
\label{Cayleysimplectic}
\eeq
\noindent
where $I_4(\Gamma)$ is the usual quartic invariant known from studies concerning the STU model
\begin{eqnarray}
I_4(\Gamma)=&-&(P^IQ_I)^2+4[(P^1Q_1)(P^2Q_2)+(P^2Q_2)(P^3Q_3)+(P^1Q_1)(P^3Q_3)]
\nonumber\\&+&4Q_0P^1P^2P^3-4P^0Q_1Q_2Q_3.
\label{i4}
\end{eqnarray}
\noindent

According to the general theory the data giving rise to BPS black hole solutions are incorporated into a  ${\cal H}(\tau)$
 subject to two constraints\cite{Sabra,Denef}
\beq
I_4(\overline{\Gamma})=-4D(\vert\overline{\Gamma}\rangle)=1,\quad
(\Gamma,\overline{\Gamma})\equiv P^I\overline{Q}_I-Q_J\overline{P}^J=0.
\label{constraints}
\eeq
\noindent
As far as these constraints hold we can completely characterize any $I_4(\Gamma)>0$ solution by generalizing the attractor equations to the so called stabilization equations\cite{Sabra}.
The warp factor of Eq.~(\ref{ansatz}) is
\beq
e^{-4U(\tau)}=I_4({\cal H}(\tau))=-4D(\vert{\cal H}(\tau)\rangle).
\label{warphyper}
\eeq

Since for normalized states the quantity 
\beq
0\leq {\tau}_{123}\equiv 4\vert D(\vert\psi\rangle)\vert\leq 1 
\label{threetangle}
\eeq
\noindent
is characterizing the tripartite entanglement of three-qubit systems\cite{Kundu} we observe that the warp factor can be regarded as a $\tau=\frac{1}{r}$ dependent entanglement measure describing the tripartite 
entanglement of a state encapsulating the details of the charge configuration and the asymptotic values for the moduli, i.e.
\beq
e^{-4U(r)}={\tau}_{123}(\vert{\cal H}(r)\rangle).
\label{w}
\eeq
\noindent
In this picture the first constraint of Eq.~(\ref{constraints}) means that
for the state of Eq.~(\ref{harmonicstate}) 
the asymptotic value of the "three-tangle" is normalized to one.

In order to find an entanglement based meaning for the second constraint of Eq.~(\ref{constraints}) notice that the $16$ quantities of the states $\vert\Gamma\rangle$ and $\vert\overline{\Gamma}\rangle$ can be organized into a {\it four-qubit state}. Indeed let us define a state $\vert\gamma\rangle$ with its $16$ amplitudes ${\gamma}_{mlkj}$ given by
\beq
{\gamma}_{0lkj}\equiv {\Gamma}_{lkj},\quad {\gamma}_{1lkj}\equiv \overline{\Gamma}_{lkj}.
\label{fourqubit}
\eeq
\noindent
then a four-qubit entanglement measure invariant under $\mathrm{SL}(2, {\mathbb C})^{\times 4}$ is given by\cite{Luque} $\vert\sigma_{1234}\vert$ where
\beq
{\sigma}_{1234}(\vert\gamma\rangle)\equiv {\gamma}_0{\gamma}_{15}-{\gamma}_1{\gamma}_{14}-{\gamma}_2{\gamma}_{13}+{\gamma}_3{\gamma}_{12}-{\gamma}_4{\gamma}_{11}+{\gamma}_5{\gamma}_{10}+
{\gamma}_6{\gamma}_9-{\gamma}_7{\gamma}_8
\label{4qubitinv}
\eeq
where for simplicity we again used the decimal labelling.
Now using the definitions as given by Eqs.~(\ref{chargegamma}) and (\ref{overgamma}) it is straightforward to check that
\beq
(\Gamma,\overline{\Gamma})=\sigma_{1234}(\vert\gamma\rangle)
\label{sigmavanish}
\eeq
\noindent
meaning that our second constraint is equivalent to the vanishing of the entanglement invariant $\sigma_{1234}$.
Hence we conclude that both of the constraints describing the BPS black hole solutions have a characteristic meaning in our entanglement based reformulation.
Recalling also that the value of ${\tau}_{123}(\vert\Gamma\rangle)$ is related to the the entropy of the BPS STU black hole\cite{Duff1,Linde,Levay1} we can summarize these results as
\beq
S=\pi\sqrt{\tau_{123}(\vert\Gamma\rangle)},
\qquad
{\tau}_{123}(\vert\overline{\Gamma}\rangle)=1,
\qquad
{\sigma}_{1234}(\vert\gamma\rangle)=0.
\label{entropybps}
\eeq
\noindent 

Now in order to set the stage for the following generalizations we review and slightly extend the known results concerning the distillation procedure for BPS solutions\cite{Levay1}. 
In the following for simplicity we consider the  $D0-D4$ system.
In this case we have $Q_0>0$ and $P^i>0$ but $P^0=Q_i=0$.
Generalizing the simple BPS solution we also include non-trivial B-fields as follows\cite{seed}. First we define our state                               $\vert\overline{\Gamma}\rangle$ with  amplitudes for Eq.~(\ref{overgamma}) as
\beq
\overline{P}^0=\frac{1}{\sqrt{2}}\sin\delta,\quad
\overline{P}^{1,2,3}=\frac{1}{\sqrt{2}}(\cos\delta +\sin\delta B^{1,2,3}),
\eeq
\beq
\overline{Q}_1=\frac{1}{\sqrt{2}}(\sin\delta[1-B^2B^3]-\cos\delta[B^1+B^2]),\quad \text{and cyclic permutations}
\eeq
\beq
\overline{Q}_0=\frac{1}{\sqrt{2}}\left[(B^1+B^2+B^3-B^1B^2B^3)\sin\delta +(1-B^1B^2-B^2B^3-B^1B^3)\cos\delta\right].
\eeq
Here $B^1,B^2,B^3$ are related to the asymptotic values for the moduli as
\beq
z^a\vert_{\tau=0}\to B^a-i,
\label{asympmod}
\eeq
i.e.~the asymptotic volume moduli are normalized, but we keep the asymptotic $B$-fields as free variables.

In terms of our three-qubit states there is a nice way of understanding this choice for $\overline{P}^I$ and  $\overline{Q}_I$  and also the meaning of the additional parameter $\delta$.
First just like in Eq.~(\ref{3qubitallapot}) we define a new moduli dependent state $\vert\overline{\Psi}(\tau)\rangle$
as
\beq
\vert\overline{\Psi}(\tau)\rangle\equiv ({\cal S}_3(\tau)\otimes{\cal S}_2(\tau)\otimes {\cal S}_1(\tau))\vert\overline{\Gamma}\rangle,
\label{overstate}
\eeq 
\noindent
where for the definition of ${\cal S}_{1,2,3}$ see Eq.~(\ref{Smatrix}).
We would like to see how this state behaves at the asymptotically Minkowski region.
A straightforward calculation using Eq.~(\ref{asympmod}) shows that
\beq
\vert\overline{\Psi}(0)\rangle=\frac{1}{\sqrt{2}}(e^{-i\delta}\vert 000\rangle-e^{i\delta}\vert 111\rangle),
\label{overghz}
\eeq
\noindent
i.e.~the parameter $\delta$ is related to the phase of a generalized GHZ state.
Notice that 
thanks to 
our inclusion of the factor $\frac{1}{\sqrt{2}}$ in the (\ref{overgamma}) definition of $\vert\overline{\Gamma}\rangle$
this state 
is {\it normalized}.
On the other hand, according to Eq.~(\ref{Norma})  a similar inclusion for the definition of the companion state $\vert\Gamma\rangle$ (Eq.~(\ref{cgamma})) has also fixed the norm of the state
$\vert\Psi\rangle$ to be the black hole potential.
Knowing that for the normalized generalized GHZ state of Eq.~(\ref{overghz}) ${\tau}_{123}=4\vert D\vert= 1$ by virtue of Eqs.~(\ref{relativeinv}), (\ref{i4})
 and (\ref{overstate})
we immediately get
$I_4(\overline{\Gamma})=1$, i.e.~the first of our constraint is satisfied. 

In our recent paper\cite{Levay2} it was shown that the attractor mechanism for STU BPS black holes
can be reinterpreted as a distillation mechanism of a GHZ state at the horizon
($\tau=\infty$) of the form
\beq
\vert\Psi(\infty)\rangle=(I_4(\Gamma))^{1/4}\frac{1}{\sqrt{2}}\Bigl(e^{-i\Delta}\vert 000\rangle 
-e^{i\Delta}\vert 111\rangle\Bigr),
\label{distill}
\eeq
\noindent
where the explicit expression of $\Delta$ is given by\cite{Levay3}
\beq
\cot\Delta=\frac{1}{P^0}\frac{\partial}{\partial Q_0}\sqrt{I_4(\Gamma)},
\label{fazis}
\eeq
\noindent 
where for the definition of $I_4(\Gamma)$ see Eq.~(\ref{i4}). 
Comparing Eqs.~(\ref{overghz}) and (\ref{distill}) we see that we have an attractor at the horizon ($\tau=\infty$) and an attractor at the asymptotically Minkowski region ($\tau =0$). These attractors can be described by the flows of the charge and moduli dependent states $\vert\Psi(\tau)\rangle$ and $\vert\overline{\Psi}(\tau)\rangle$ respectively, each of them producing a maximally entangled GHZ state at the attractor points defined by $\tau=\infty$ and $\tau =0$.
The attractor at $\tau=0$ giving rise to the GHZ-state of Eq.~(\ref{overghz}) is the well-known attractor at infinity\cite{infti}, a map between the $6$ real moduli and the $8$ constants in the harmonic functions subject to $2$ constraints. 
We can summarize these considerations by noting 
that for the relevant flows and attractor points we have
\beq
\vert\vert\overline{\Psi}(0)\vert\vert^2=1,\qquad
\vert\vert{\Psi}(\infty)\vert\vert^2= V_{BH}(\infty)=\sqrt{I_4(\Gamma)}.
\label{compare1}
\eeq
\noindent

Now we compare the flows $\vert\overline{\Psi}(\tau)\rangle$
and $\vert\Psi(\tau)\rangle$ by calculating
\beq
\langle\overline{\Psi}(\tau)\vert 
\sigma_3\otimes\sigma_3\otimes\sigma_3\vert
\Psi(\tau)\rangle.
\label{phaseflip}
\eeq
\noindent
This quantity is the transition amplitude between the phase-flipped state
$(\sigma_3\otimes \sigma_3\otimes\sigma_3)\vert\Psi(\tau)\rangle$ and the one
$\vert\overline{\Psi}(\tau)\rangle$.
Using Eqs.~(\ref{3qubitallapot}) and (\ref{overstate})
and the fact that 
${\cal U}^{\dagger}\sigma_3{\cal U}=-\sigma_2$  and $S^T\sigma_2S=\sigma_2$ for $S\in \mathrm{SL}(2,\mathbb R)$, we get
\beq
\langle\overline{\Psi}(\tau)\vert \sigma_3\otimes \sigma_3\otimes\sigma_3\vert
\Psi(\tau)\rangle=-\langle\overline{\Gamma}\vert \sigma_2\otimes \sigma_2\otimes \sigma_2\vert\Gamma\rangle=i(\overline{\Gamma},\Gamma)=i{\sigma}_{1234}(\vert\gamma\rangle)=0.
\label{vanish}
\eeq
\noindent
This shows that the phase (sign)-flipped version of $\Psi(\tau)$ is always orthogonal to the companion state $\overline{\Psi}(\tau)$.
It is also clear that this amplitude is of purely topological in origin.
(E.g. in the type IIB duality frame it is related to the intersection product\cite{Denef} on $T^6$.)
Due to the fact that $H\sigma_3H=\sigma_1$, where $\sigma_1$ is the bit flip operator we can alternatively conclude that the bit flipped version of the Fourier transform of one of the  states is orthogonal to the Fourier transform of the other.
Recall also that according to Eq.~(\ref{flipcov}) such bit flip errors are related to the action of the flat  covariant derivatives with respect to the moduli.

As a last application of the vanishing of the amplitude of Eq.~(\ref{vanish}),
Eqs.~(\ref{overghz}) and (\ref{centralcharge}) give the meaning of the parameter $\delta$ as the phase of the central charge $Z$ (up to a shift by $\pi$).
These results shed some light on the quantum information theoretic meaning of   the
second constraint
of Eq.~(\ref{constraints}).

In closing this section we present the well-known solution of the stabilization equations giving the moduli fields as a function of $\tau$ \cite{Sabra,Denef,seed}
\beq
z^1=\frac{-H_1H^1+H_0H^0+H_2H_2+H_3H^3-ie^{-2U}}{2(H^2H^3-H^0H_1)},
\label{solutionbps}
\eeq
\noindent
with the remainig equations for $z^2$ and $z^3$ are arising after cyclic permutation of the numbers $1,2,3$. Here $H^I=\overline{P}^I+P^I\tau$, $H_J=\overline{Q}_J+Q_J\tau$, with $I,J=0,1,2,3$, and the warp factor is given by Eq.~(\ref{warphyper}). 
It is straightforward to check that the solution $z^a(\tau)$ satisfies the correct asymptotic behavior of Eq.~(\ref{asympmod}) and 
\beq
z^a(\infty)=-i\sqrt{\frac{2Q_0P^a}{s_{abc}P^bP^c}},\qquad s_{abc}\equiv \vert{\varepsilon}_{abc}\vert.
\label{attractorvalue}
\eeq
\noindent
By virtue of Eqs.~(\ref{000})-(\ref{conj}) and (\ref{asympmod})
it is clear that the asymptotic values of the $8$ amplitudes of $\vert\Psi(0)\rangle$ are generally non zero. However,  the flow $z^a(\tau)$ in moduli space giving rise to the flow $\vert\Psi(\tau)\rangle$
results in a GHZ state at the horizon having merely $2$ nonvanishing amplitudes
(see Eq.~(\ref{distill})).

This distillation process is the one that was studied for BPS solutions\cite{Levay1}. 
Later a non-BPS generalization was also given\cite{Levay3}. In this case the distillation process gives rise to graph states\cite{graph} at the horizon. However, the analysis in these papers was restricted merely to double extremal solutions\cite{Behrndt} for which the moduli are constant even away from the horizon.
Hence in these studies the important question of how the distillation process 
becomes unfolded as $\tau$  changes have not been addressed. 
Our aim in the next section is to investigate such issues by studying the non-BPS flow explicitly.

\section{The flow for the non-BPS D0-D4 system}
\label{secSeedFlow}

In this section we study the flow $\vert\Psi(\tau)\rangle$ answering the full radial flow $z^a(\tau)$ obtained for the $5$ parameter family 
of non-BPS seed solutions of a $D0-D4$ system discribed by Gimon et.al.\cite{seed}
More precisely it turns out to be rewarding to study the properties of a related flow $\vert\chi(\tau)\rangle$ instead by multiplying $\vert\Psi(\tau)\rangle$ by the warp factor.
Then we show that the Fourier amplitudes of this new state satisfy a first order system of differential equations in accordance with our expectation coming from previous studies\cite{fake,Pioline,Stelle}.

First we address the quantum information theoretic aspects of the seed solution.
For the $D0-D4$ system we chose $Q_0<0$ and $P^a>0$, $a=1,2,3$.
Let us define
\beq
z^a=x^a-iy^a,\qquad x^a=R_at_a,\quad y^a=R_ae^{\phi_a},\quad R_a=\sqrt{\frac{-2Q_0P^a}{s_{abc}P^bP^c}}.
\label{jelolesek}
\eeq
\noindent
From Eq.~(\ref{Euler}) the equations to be solved are
\beq
\frac{\dd}{\dd\tau}\left(\dot{t}_ae^{-2\phi_a}\right)=2e^{2U}\frac{\partial V_{BH}}{
\partial t_a},\qquad a=1,2,3,
\label{egy}
\eeq
\noindent
\beq
\ddot{\phi}_a+\left(\dot{t}_ae^{-\phi_a}\right)^2=2e^{2U}\frac{\partial V_{BH}}
{\partial \phi_a},\qquad a=1,2,3,
\label{ketto}
\eeq
\noindent
\beq
\ddot{U}=e^{2U}V_{BH}.
\label{harom}
\eeq
\noindent
Moreover, according to Eq.~(\ref{constraint}) we also have the constraint
\beq
\dot{U}^2+\frac{1}{4}\sum_a\left(\dot{\phi}_a^2+[\dot{t}_ae^{-\phi_a}]^2\right)=e^{2U}V_{BH}.
\label{negy}
\eeq
\noindent
As a first step we define a new three-qubit state by incorporating also the warp factor as
\beq
\vert\chi(\tau)\rangle\equiv e^{U(\tau)}\vert\Psi(\tau)\rangle=e^{U(\tau)}({\cal S}_3\otimes{\cal S}_2\otimes {\cal S}_1)\vert\Gamma\rangle.
\label{universalstate}
\eeq
\noindent
This state depends on the charges, the moduli {\it and} the warp factor.
The discrete Fourier transform of this state (for the definitions see Eqs.~(\ref{Smatrix}), (\ref{Hadamard}), (\ref{Fourier}) and (\ref{phasegate}) )
\beq
\vert\tilde{\chi}(\tau)\rangle=(H\otimes H\otimes H)\vert \chi(\tau)\rangle=e^{U(\tau)}({\cal P}\otimes{\cal P}\otimes{\cal P})(S_3\otimes S_2\otimes S_1)\vert\Gamma\rangle,
\label{unifourier}
\eeq
\noindent
will play a particularly important role in the following.

For later use we write out explictly the amplitudes of this Fourier transformed state $\vert\tilde{\chi}(\tau)\rangle$
\beq
\tilde{\chi}_{001}=\frac{1}{2}\vert I_4\vert^{\frac{1}{4}}e^{\beta+\phi_2+\phi_3},
\qquad \tilde{\chi}_{010}=\frac{1}{2}\vert I_4\vert^{\frac{1}{4}}e^{\beta+\phi_1+\phi_3},
\qquad \tilde{\chi}_{100}=\frac{1}{2}\vert I_4\vert^{\frac{1}{4}}e^{\beta+\phi_1+\phi_2},
\label{amp124}
\eeq
\noindent
\beq
i\tilde{\chi}_{110}=\frac{1}{2}\vert I_4\vert^{\frac{1}{4}}e^{\beta+\phi_1}(t_2+t_3),\quad
i\tilde{\chi}_{101}=\frac{1}{2}\vert I_4\vert^{\frac{1}{4}}e^{\beta+\phi_2}(t_1+t_3),\quad
i\tilde{\chi}_{011}=\frac{1}{2}\vert I_4\vert^{\frac{1}{4}}e^{\beta+\phi_3}(t_1+t_2),
\label{ampl653}
\eeq
\noindent
\beq
\tilde{\chi}_{000}=0,\qquad \tilde{\chi}_{111}=-\frac{1}{2}\vert I_4\vert^{\frac{1}{4}}e^{\beta}(1+t_1t_2+t_2t_3+t_3t_1).
\label{ampl07}
\eeq
\noindent
Here 
\beq
\beta=U-\frac{1}{2}(\phi_1+\phi_2+\phi_2),\qquad \vert I_4\vert= -I_4=-4Q_0P^1P^2P^3>0.
\label{jel2}
\eeq
\noindent
Now in terms of these amplitudes the equations to be solved can be written as
\beq
\left(\frac{\dd}{\dd\tau}-\dot{\phi}_a\right)\left(\dot{t}_ae^{-\phi_a}\right)=-2
\langle\tilde{\chi}\vert Y_a\vert\tilde{\chi}\rangle,
\label{egyv}
\eeq
\noindent
\beq
 \ddot{\phi}_a+\left(\dot{t}_ae^{-\phi_a}\right)^2=2\langle\tilde{\chi}\vert {\cal Z}_a\vert\tilde{\chi}\rangle,
\label{kettov}
\eeq
\noindent
\beq
\ddot{U}=\langle\tilde{\chi}\vert\tilde{\chi}\rangle,
\label{haromv}
\eeq
\noindent
where the operators ${\cal Z}_1\equiv I\otimes I\otimes \sigma_3$, ${\cal Z}_2\equiv I\otimes \sigma_3\otimes I$ and ${\cal Z}_3=\sigma_3\otimes I\otimes I$
describe phase flips of the $a$th qubit in the Fourier transformed base.
The quantities $Y_a$ are defined using $\sigma_2$ accordingly.
Notice also that the operators $\frac{1}{2}({\bf 1}+{\cal Z}_a)$ where ${\bf 1}$ is the $8\times 8$ unit matrix, are projection operators.
Hence by adding the {\it half} of Eqs.~(\ref{kettov}) to Eq.~(\ref{haromv})
gives rise to three equations which contain merely {\it three amplitudes}
($\tilde{\chi}_0=0$) on the right hand side.
These manipulations also justify the introduction of a new variable\cite{seed,Kubota}
\beq
{\alpha}_a\equiv U+\frac{1}{2}\phi_a,
\label{alpha}
\eeq
\noindent
Using this new variable instead of Eqs.~(\ref{kettov}) we can use the ones
\beq
\ddot{\alpha}_1-2\left(\frac{i}{2}\dot{t}_ae^{-\phi_a}\right)^2=2(\tilde{\chi}_2^2+
\tilde{\chi}_4^2-\tilde{\chi}_6^2),
\label{1v}
\eeq
\noindent
\beq
\ddot{\alpha}_2-2\left(\frac{i}{2}\dot{t}_2e^{-\phi_2}\right)^2=2(\tilde{\chi}_1^2+
\tilde{\chi}_4^2-\tilde{\chi}_5^2),
\label{2v}
\eeq
\noindent
\beq
\ddot{\alpha}_3-2\left(\frac{i}{2}\dot{t}_3e^{-\phi_3}\right)^2=2(\tilde{\chi}_1^2+
\tilde{\chi}_2^2-\tilde{\chi}_3^2),
\label{3v}
\eeq
\noindent
where from now on we use decimal labelling for our amplitudes.
For Eq.~(\ref{egyv}) we have the form
\beq
\begin{pmatrix}\frac{\dd}{\dd\tau}-\dot{\phi}_1&0&0\\0&\frac{\dd}{\dd\tau}-\dot{\phi}_2&0\\0&0&\frac{\dd}{\dd\tau}-\dot{\phi}_3    \end{pmatrix}
\begin{pmatrix}\frac{i}{2}\dot{t}_1e^{-\phi_1}\\
\frac{i}{2}\dot{t}_2e^{-\phi_2}\\\frac{i}{2}\dot{t}_3e^{-\phi_3}\end{pmatrix}=2
\begin{pmatrix}\tilde{\chi}_7&-\tilde{\chi}_4&-\tilde{\chi}_2\\
-\tilde{\chi}_4&\tilde{\chi}_7&-\tilde{\chi}_1\\ -\tilde{\chi}_2&-\tilde{\chi}_1&\tilde{\chi}_7\end{pmatrix}
\begin{pmatrix}\tilde{\chi}_6\\
\tilde{\chi}_5\\\tilde{\chi}_3\end{pmatrix},
\label{egyvv}
\eeq
\noindent
and the sum of Eq.~(\ref{haromv}) and $-\frac{1}{2}$ the sum of Eqs.~(\ref{kettov}) gives
\beq
\ddot{\beta}+2\sum_a\left(\frac{i}{2}\dot{t}_ae^{-\phi_a}\right)^2=4\tilde{\chi}_7^2-2(\tilde{\chi}_6^2+\tilde{\chi}_5^2+\tilde{\chi}_3^2).
\label{haromvv}
\eeq
\noindent
Now, the crucial observation which enables an explicit construction of the seed solutions is the fact that in the Fourier transformed basis we have {\it seven} nonvanishing amplitudes, and the constraint Eq.~(\ref{negy})
is consisting of squares of seven terms.
Using the decimal labelling for the amplitudes of $\vert\tilde{\chi}\rangle$ this constraint can be written in the form
\beq
\dot{\beta}^2+\sum_{a\neq b\neq c}(\dot{\alpha}_a+\dot{\alpha}_b-\dot{\alpha}_c)^2+\sum_a(e^{-\phi_a}\dot{t}_a)^2=4({\chi}_7^2+{\chi}_1^2+{\chi}_2^2+{\chi}_4^2-{\chi}_6^2-{\chi}_5^2-{\chi}_3^2).
\eeq
A natural choice to satisfy this constraint up to a sign is 
\beq
\pm\tilde{\chi}_7=\frac{1}{2}\dot{\beta},\qquad
\pm\tilde{\chi}_6=\frac{i}{2}e^{-\phi_1}\dot{t}_1,\quad
\pm\tilde{\chi}_5=\frac{i}{2}e^{-\phi_2}\dot{t}_2,\quad
\pm\tilde{\chi}_3=\frac{i}{2}e^{-\phi_3}\dot{t}_3,
\label{1}
\eeq
\noindent
\beq
\pm\tilde{\chi}_1=\frac{1}{2}(\dot{\alpha}_1-\dot{\alpha}_2-\dot{\alpha}_3),\quad
\pm\tilde{\chi}_2=\frac{1}{2}(\dot{\alpha}_2-\dot{\alpha}_1-\dot{\alpha}_3),\quad
\pm\tilde{\chi}_4=\frac{1}{2}(\dot{\alpha}_3-\dot{\alpha}_1-\dot{\alpha}_2).
\label{2}
\eeq
\noindent
Comparing these with the explicit form of the Fourier amplitudes of Eqs.~(\ref{amp124})-(\ref{ampl07}) with $\tilde{\chi}_{1,2,4}$ rewritten as
\beq
\tilde{\chi}_1=\frac{1}{2}\vert I_4\vert^{1/4}e^{\alpha_2+\alpha_3-\alpha_1},\quad
\tilde{\chi}_2=\frac{1}{2}\vert I_4\vert^{1/4}e^{\alpha_3+\alpha_1-\alpha_2},\quad
\tilde{\chi}_4=\frac{1}{2}\vert I_4\vert^{1/4}e^{\alpha_1+\alpha_2-\alpha_3},
\label{newampl124}
\eeq
\noindent
we get the following set of {\it first order differential equations}
\beq
\dot{t}_a=\mp \vert I_4\vert^{1/4}(t_b+t_c)e^{3\alpha_a-\alpha_b-\alpha_c},
\label{1e}
\eeq
\noindent
\beq
\dot{\beta}=\mp \vert I_4\vert^{1/4}e^{\beta}(1+t_1t_2+t_2t_3+t_3t_1),
\label{2e}
\eeq
\noindent
\beq
(\dot{\alpha}_a+\dot{\alpha}_b-\dot{\alpha}_c)=\mp \vert I_4\vert^{1/4}e^{\alpha_a+\alpha_b-\alpha_c},
\label{3e}
\eeq
\noindent
where $a\neq b\neq c$.
The solutions to these equations with the {\it upper sign} choice were given in Gimon et.al.\cite{seed} Before recalling these solutions we show that these choices automatically solve Eqs.~(\ref{1v})-(\ref{haromvv}). 
Let us substitute instead of $\frac{i}{2}\dot{t}_ae^{-\phi_a}$ occurring in these equations the Fourier amplitudes of  Eqs.~(\ref{1}).
Then we get
\beq
\ddot{\beta}=4(\tilde{\chi}_7^2-\tilde{\chi}_6^2-\tilde{\chi}_5^2-\tilde{\chi}_3^2),
\label{11}
\eeq
\noindent
\beq
\ddot{\alpha}_1=2(\tilde{\chi}_2^2+\tilde{\chi}_4^2),\quad
\ddot{\alpha}_2=2(\tilde{\chi}_1^2+\tilde{\chi}_4^2),\quad
\ddot{\alpha}_3=2(\tilde{\chi}_1^2+\tilde{\chi}_2^2).
\label{12}
\eeq
\noindent
In Eq.~(\ref{egyvv}) ve also replace $\tilde{\chi}_7$ by $\frac{1}{2}\dot{\beta}$ (we have chosen the upper sign) to get
\beq
\frac{\dd}{\dd\tau}\begin{pmatrix}\tilde{\chi}_6\\\tilde{\chi}_5\\\tilde{\chi}_3\end{pmatrix}=2\begin{pmatrix}\frac{1}{2}(\dot{\beta}+\dot{\phi}_1)&-\tilde{\chi}_4&-\tilde{\chi}_2\\-\tilde{\chi}_4&\frac{1}{2}(\dot{\beta}+\dot{\phi}_2)&-\tilde{\chi}_1\\-\tilde{\chi}_2&-\tilde{\chi}_1&\frac{1}{2}(\dot{\beta}+\dot{\phi}_3)\end{pmatrix}
\begin{pmatrix}\tilde{\chi}_6\\\tilde{\chi}_5\\\tilde{\chi}_3\end{pmatrix}.
\label{13}
\eeq
\noindent
Now using the explicit expressions for the Fourier amplitudes it is easy to check that these equations are indeed satisfied.

An interesting possibility is to write down these equations as first order equations for the Fourier amplitudes.
From Eqs.~(\ref{11}) and (\ref{13}) we get
\beq
\frac{\dd}{\dd\tau}\begin{pmatrix}\tilde{\chi}_7\\\tilde{\chi}_6\\\tilde{\chi}_5\\\tilde{\chi}_3\end
{pmatrix}=2\begin{pmatrix}\frac{1}{2}\dot{\beta}&-\tilde{\chi}_6&-\tilde{\chi}_5& -\tilde{\chi}_3\\0&\frac{1}{2}(\dot{\beta}+\dot{\phi}_1)&-\tilde{\chi}_4&
-\tilde{\chi}_2\\0&-\tilde{\chi}_4&\frac{1}{2}(\dot{\beta}+\dot{\phi}_2)&         -\tilde{\chi}_1
\\0&-\tilde{\chi}_2&-\tilde{\chi}_1&\frac{1}{2}(\dot{\beta}+\dot{\phi}_3)         \end{pmatrix}
\begin{pmatrix}\tilde{\chi}_7\\\tilde{\chi}_6\\\tilde{\chi}_5\\\tilde{\chi}_3\end{pmatrix}.
\label{E1}
\eeq
\noindent
Similarly from Eq.~(\ref{12}) using Eq.~(\ref{newampl124})
the corresponding equation is
\beq
\frac{\dd}{\dd\tau}\begin{pmatrix}\tilde{\chi}_1\\\tilde{\chi}_2\\\tilde{\chi}_4\end{pmatrix}=2\begin{pmatrix}-\tilde{\chi}_1&0&0\\0&-\tilde{\chi}_2&0\\0&0&-\tilde{\chi}_4\end{pmatrix}
\begin{pmatrix}\tilde{\chi}_1\\\tilde{\chi}_2\\\tilde{\chi}_4\end{pmatrix}.
\label{E2}
\eeq
\noindent
An alternative form for Eq.~(\ref{E1}) can be given by noticing that
\beq
\dot{\beta}+\dot{\phi_a}=2\dot{U}-(\dot{\alpha}_b+\dot{\alpha}_b-\dot{\alpha_a})=2\dot{U}+\vert I_4\vert^{1/4}e^{\alpha_b+\alpha_c-\alpha_a}=2\dot{U}+2\tilde{\chi}_{1,2,4},
\eeq
hence we have
\beq
\frac{\dd}{\dd\tau}\begin{pmatrix}\tilde{\chi}_7\\\tilde{\chi}_6\\\tilde{\chi}_5\\
\tilde{\chi}_3\end
{pmatrix}=2\begin{pmatrix}\tilde{\chi}_7&-\tilde{\chi}_6&-\tilde{\chi}_5&
-\tilde{\chi}_3\\0&\dot{U}+\tilde{\chi}_1&-\tilde{\chi}_4&
-\tilde{\chi}_2\\0&-\tilde{\chi}_4&\dot{U}+\tilde{\chi}_2&
-\tilde{\chi}_1
\\0&-\tilde{\chi}_2&-\tilde{\chi}_1&\dot{U}+\tilde{\chi}_4
\end{pmatrix}
\begin{pmatrix}\tilde{\chi}_7\\\tilde{\chi}_6\\\tilde{\chi}_5\\\tilde{\chi}_3   \end{pmatrix}.
\label{E3}
\eeq
\noindent
Eqs.~(\ref{E2}) and (\ref{E3}) are first order equations containing
the Fourier amplitudes of our charge, moduli and warp factor dependent states
{\it and} $\dot{U}$.
However, we can express $\dot{U}$ in terms of some of the $\tilde{\chi}$ s
as follows.
Let us define
\beq
w\equiv \frac{1}{2}(\tilde{\chi}_7-\tilde{\chi}_1-\tilde{\chi}_2-\tilde{\chi}_4).
\label{fakewarp}
\eeq
\noindent
Then from Eqs.~(\ref{E2}) and (\ref{E3}) we get
\beq
\frac{\dd}{\dd\tau}w=(\tilde{\chi}_1)^2+(\tilde{\chi}_2)^2+(\tilde{\chi}_4)^2+(\tilde{\chi}_7)^2-(\tilde{\chi}_6)^2-(\tilde{\chi}_5)^2-(\tilde{\chi}_3)^2=e^{2U}V_{BH} .
\label{fakeeq}
\eeq
\noindent
Hence fom Eq.~(\ref{haromv})
\beq
\frac{\dd}{\dd\tau}U=w\equiv e^{U}{\cal W}.
\label{fake}
\eeq
\noindent
The new quantity ${\cal W}$
\beq
{\cal W}=\frac{1}{2}(\tilde{\Psi}_7-\tilde{\Psi}_1-\tilde{\Psi}_2-\tilde{\Psi}_4)=
\frac{1}{2}(\tilde{\Psi}_{111}-\tilde{\Psi}_{001}-\tilde{\Psi}_{010}-\tilde{\Psi}_{100}),
\label{explfake}
\eeq
\noindent
is the fake superpotential\cite{fake,stu} where it is easy to check that its explicit form coincides with the negative of the one as given by Eq.~(6.8) of the paper
by Bellucci et.al.\cite{stu}
Notice also that the fake superpotential contains only Fourier amplitudes
of {\it odd parity} of the charge and moduli dependent three-qubit state $\vert\Psi\rangle$. 
This will be of some importance in the next section.

Since we have $\dot{U}=\frac{1}{2}(\tilde{\chi}_7-\tilde{\chi}_1-\tilde{\chi}_2-\tilde{\chi}_4)=w$ we can write Eq.~(\ref{E3}) in the final form 
\beq
\frac{\dd}{\dd\tau}\begin{pmatrix}\tilde{\chi}_7\\\tilde{\chi}_6\\\tilde{\chi}_5\\
\tilde{\chi}_3\end
{pmatrix}=2\begin{pmatrix}\tilde{\chi}_7&-\tilde{\chi}_6&-\tilde{\chi}_5&
-\tilde{\chi}_3\\0&w+\tilde{\chi}_1&-\tilde{\chi}_4&
-\tilde{\chi}_2\\0&-\tilde{\chi}_4&w+\tilde{\chi}_2&
-\tilde{\chi}_1
\\0&-\tilde{\chi}_2&-\tilde{\chi}_1&w+\tilde{\chi}_4
\end{pmatrix}
\begin{pmatrix}\tilde{\chi}_7\\\tilde{\chi}_6\\\tilde{\chi}_5\\\tilde{\chi}_3
\end{pmatrix}.
\label{E4}
\eeq
\noindent
Hence Eqs.~(\ref{E2}) and (\ref{E4}) show that in the case of the seed solution for the non-BPS $Z\neq 0$  $D0-D4$ system the $\tau$ derivatives of the Fourier
amplitudes of our {\it charge, moduli and warp factor}-dependent state $\vert\chi\rangle$ can be expressed entirely in terms of the Fourier amplitudes.

\section{The attractor mechanism as a distillation procedure}
\label{secSeedDist}

In this section we would like to demonstrate how the radial flow studied in the previous section gives rise to the distillation of a special three-qubit state at the black hole horizon.
In order to see this procedure unfolding all we have to do is to use the solutions of the first order equations Eqs.~(\ref{1e})-(\ref{3e}) to obtain 
explicit expressions for the Fourier amplitudes $\tilde{\chi}_{lkj}(\tau)$.
It means that starting from the asymptotic values $\tilde{\chi}_{lkj}(0)$ in the region with Minkowski geometry at the limit $\tau\to \infty$ we obtain the ones $\tilde{\chi}_{lkj}(\infty)$ at the horizon with $adS_2\times S^2$ geometry.
The solutions of Eqs.~(\ref{1e})-(\ref{3e}) are \cite{seed}
\beq
e^{{\alpha}_a+{\alpha}_b-{\alpha}_c}=\frac{1}{d_c+\vert I_4\vert^{1/4}\tau}\equiv\frac{1}{\hat{h}_c},\qquad a\neq b\neq c,\quad a,b,c=1,2,3,
\label{al}
\eeq
\noindent
\beq
t_a=\frac{B}{\hat{h}_b\hat{h}_c},\qquad e^{-\beta}=-\hat{h}_0-\frac{B^2}{\hat{h}_1\hat{h}_2\hat{h}_3},\qquad \hat{h}_0\equiv -d_0-\vert I_4\vert^{1/4}\tau
\label{tbeta}
\eeq
\noindent
where
\beq
d_a=\frac{\vert I_4\vert^{1/4}}{\sqrt{2}P^a},\qquad d_0=-\frac{\vert I_4\vert^{1/4}}{\sqrt{2}Q_0}(1+B^2)
\label{boundaryvalues}
\eeq
\noindent
where for the definition of $\vert I_4\vert$ see Eq.~(\ref{jel2}).
Notice that according to Eq.~(\ref{normmagnetic}) and (\ref{jelolesek}) for this $5$ parameter  solution we have
\beq
z^a=R_a\frac{B-ie^{-2U}}{\frac{1}{2}s_{abc}\hat{h}_b\hat{h}_c},\qquad e^{-4U}=
-\hat{h}_0\hat{h}_1\hat{h}_2\hat{h}_3-B^2
\label{modulimegoldas}
\eeq
\noindent
hence
$B\equiv x^1(0)=x^2(0)=x^3(0)$. 

In order to write down the explicit form of the amplitudes of our charge, moduli and warp factor dependent Fourier transformed state it is useful to introduce the new harmonic functions
\beq
H^a(\tau)=\frac{1}{\sqrt{2}}+P^a\tau=\frac{P^a}{\vert I_4\vert^{1/4}}\hat{h}_a(\tau),\quad
H_0(\tau)=-\frac{1}{\sqrt{2}}(1+B^2)+Q_0\tau=-\frac{Q_0}{\vert I_4\vert^{1/4}}\hat{h}_0,
\label{nwharmonic}
\eeq
\noindent
and the warp factor
\beq
e^{-4U(\tau)}=-4H_0(\tau)H^1(\tau)H^2(\tau)H^3(\tau)-B^2.
\label{physicalwarp}
\eeq
\noindent
Using these results we get the following results for the Fourier amplitudes
for our charge, moduli and warp factor dependent state $\vert\tilde{\chi}\rangle$
\beq
\tilde{\chi}_{000}(\tau)=0,\qquad
\tilde{\chi}_{001}(\tau)=\frac{P^1}{2H^1(\tau)},\qquad 
\tilde{\chi}_{010}(\tau)=\frac{P^2}{2H^2(\tau)},\qquad
\tilde{\chi}_{100}(\tau)=\frac{P^3}{2H^3(\tau)},
\label{124fin}
\eeq
\noindent
\beq
\tilde{\chi}_{110}(\tau)=-\frac{i}{2}e^{2U(\tau)}\left[\frac{P^2}{H^2(\tau)}-
\frac{P^3}{H^3(\tau)}\right],
\label{110fin}
\eeq
\noindent
\beq
\tilde{\chi}_{101}(\tau)=-\frac{i}{2}e^{2U(\tau)}\left[\frac{P^1}{H^1(\tau)}-
\frac{P^3}{H^3(\tau)}\right],
\label{101fin}
\eeq
\noindent
\beq
\tilde{\chi}_{011}(\tau)=-\frac{i}{2}e^{2U(\tau)}\left[\frac{P^1}{H^1(\tau)}-
\frac{P^2}{H^2(\tau)}\right],
\label{011fin}
\eeq
\noindent
\beq
\tilde{\chi}_{111}(\tau)=\frac{1}{2}e^{4U(\tau)}\left[4Q_0H^1(\tau)H^2(\tau)H^3(\tau)-B^2\sum_{a=1}^3\frac{P^a}{H^a(\tau)}\right].
\label{111fin}
\eeq
\noindent
From this the components of the charge and moduli dependent 
Fourier transformed state $\vert\tilde{\Psi}(\tau)\rangle$ are
\beq
\tilde{\Psi}_{lkj}(\tau)=e^{-U(\tau)}\tilde{\chi}_{lkj}(\tau).
\label{transformwarp}
\eeq
\noindent
Since
\beq
\vert\tilde{\Psi}(\tau)\rangle=(H\otimes H\otimes H)\vert\Psi(\tau)\rangle
=\sum_{lkj=0}^1\tilde{\Psi}(\tau)\vert lkj\rangle,
\label{tisztazas}
\eeq
\noindent
one can show that in the asymptotically Minkowski region we have
\begin{eqnarray}
\lim_{\tau\to 0}\vert\tilde{\Psi}(\tau)\rangle=\frac{1}{\sqrt{2}}\Bigl(P^1\vert 001\rangle &+&
P^2\vert 010\rangle +P^3\vert 100\rangle
-iB(P^2+P^3)\vert 110\rangle \nonumber\\
 &-&iB(P^1+P^3)\vert 101\rangle -iB(P^1+P^2)\vert 011\rangle\nonumber\\
&+&[Q_0-B^2(P^1+P^2+P^3)]\vert 111\rangle\Bigr).
\label{Minkowski}
\end{eqnarray}
\noindent
On the other hand at the horizon we have
\beq
\lim_{\tau\to\infty}\vert\tilde{\Psi}(\tau)\rangle=\vert I_4\vert^{1/4}\frac{1}{2}\Bigl(\vert 001\rangle +
\vert 010\rangle +\vert 100\rangle - \vert 111\rangle\Bigr).
\label{attractor}
\eeq
\noindent
This result shows that if we "start" asymptotically with the state of Eq.~(\ref{Minkowski}) with seven nonvanishing generically different amplitudes we end up with the state of Eq.~(\ref{attractor}) having merely four nonvanishing amplitudes that are the same up to a sign.
Notice also that the four nonvanishing amplitudes are having states with {\it odd parity}. This reminds us of the structure of the fake superpotential
Eq.~(\ref{explfake}).
This is as it should be since we know\cite{stu} that in the near horizon limit
${\cal W}^2$ should  give the square root of $-I_4>0$.
Indeed, since we have
\beq
\lim_{\tau\to\infty}{\tilde{\Psi}}_{1,2,4}(\tau)
=-\lim_{\tau\to\infty}\tilde{\Psi}_7(\tau)=\frac{1}{2}\vert I_4\vert^{1/4},
\label{explfake2}
\eeq
\noindent
$\lim_{\tau\to\infty}{\cal W}^2(\tau)=\vert I_4\vert^{1/2}=\sqrt{-4Q_0P^1P^2P^3}$ which is the correct value. 
According to Eqs.~(\ref{Cayleysimplectic}) and (\ref{threetangle}) in our three-qubit interpretation 
\beq
{\cal W}^2(\infty)=\sqrt{\tau_{123}(\vert\tilde{\Psi}(\infty)\rangle)},
\eeq
\noindent
i.e.~the square of the fake superpotential on the horizon is just the entanglement measure of the state of Eq.~(\ref{attractor}).
It is important to realize that the components of the fake superpotential are precisely those amplitudes of our 
$3$-qubit state which are {\it not dying out} as we are approaching the horizon. According to our previous results this also works for the BPS case
(and as will be seen in the next section even for the non-BPS case with vanishing central charge).
In order to verify this just recall that for BPS solutions $\vert Z\vert$ plays the role of the superpotential which is according to Eqs.~(\ref{000}) and (\ref{distill})
is again related to the amplitude which is not dying out in the attractor limit.

\section{Entangled states of GHZ type on the horizon}
\label{secPsi}

In the previous sections we have studied the distillation process in the special case of the non-BPS $Z\neq 0$ seed solution.
Clearly similar results can be obtained for the most general non-BPS solutions\cite{stu} 
and the non-BPS ones with vanishing central charge\cite{Scherbakov}.
In this section however, our main concern will be to present 
the explicit forms of our $3$-qubit state $\vert\Psi(\tau)\rangle$ at the horizon.
Of course the states we expect to show up are again GHZ-like states, 
but the new subtlety worth investigating in this context is the appearance of flat directions\cite{flat}.
As we have mentioned we have to make distinction between three different cases
$\frac{1}{2}$-BPS, non-BPS $Z=0$, and non-BPS $Z\neq0$ solutions.

In this section we use the $p^I$, $q_I$ quantized charges instead of the $P^I$, $Q_I$ dressed ones.
This is because we would like to use the most general non-BPS $Z\neq0$ solution\cite{stu} which has been produced by using an U-duality transformation
acting nicely on such quantized charges.
The dressed charges are rescaled quantities related to the quantized (undressed) ones
via factors coming from the asymptotic volume moduli.
Calculating the moduli $\tilde{z}^a$ using the (\ref{potential}) black hole potential with $p^I$, $q_I$,
the asymptotic volume of the tori are nontrivial $\tilde{y}^a(0)=v_a$.
In order to get $y^a(0)=1$ as in Eq.~(\ref{normmagnetic})
we have to rescale the charges with real positive dressing factors.
(For the definitions of these factors see the paper of  Gimon \textit{et.al.}\cite{seed})
The dictionary between the two conventions, i.e.~$p^I$, $q_I$ with $\tilde{z}^a=z^av_a$
and $P^I$, $Q_I$ with $z^a$ is then effected by the correspondence 
\begin{equation}
V_{BH}(P^I,Q_I,z^a)=G_N V_{BH}(p^I,q_I,\tilde{z}^a)
\end{equation}
where $G_N$ is the $D=4$ Newton constant.
The quartic invariant defined in Eq.~(\ref{i4}) can also be written in terms of the quantized charges, hence
\begin{equation}
I_4(P^I,Q_I)=G_N^2I_4(p^I,q_I).
\end{equation}
In this section we denote $I_4=I_4(p^I,q_I)$.
One can check that not only the norm of $\Psi$ i.e.~the square root of the  black hole potential,
but also our three-qubit state scales simply with $\sqrt{G_N}$
\begin{equation}
\vert\Psi(P^I,Q_I,z^a)\rangle = \sqrt{G_N}\vert\Psi(p^I,q_I,\tilde{z}^a)\rangle.
\end{equation}

\subsection{The BPS case}

After these techniqualities first we turn once again to the BPS solutions.
The black hole charge configurations supporting the $\frac{1}{2}$-BPS attractors at the event horizon 
are the ones satisfying the following set of constraints\cite{stu}
\begin{equation}
I_4>0,\qquad
p^ap^b-p^0q_c>0.
\label{bpsfeltetelek}
\end{equation}
In this case the general $\frac{1}{2}$-BPS attractor flow solution is\cite{Sabra,stu}
\begin{align}
\exp(-4U) &= I_4(h^I,h_I),\\
\tilde{x}^a(\tau) &=\frac{h^I h_I - 2 h^ah_a}{2(h^bh^c-h^0h_a)},\\
\tilde{y}^a(\tau) &=\frac{\sqrt{I_4(h^I,h_I)}}{2(h^bh^c-h^0h_a)}.
\end{align}
Here the indices $a,b,c$ are \textit{distinct} elements of the set $\{1,2,3\}$, and no summation is implied on them, on the other hand for the indices $I=0,1,2,3$ summation is understood. 
The undressed harmonic functions are defined similarly to the (\ref{Harmonic}) dressed ones
\begin{equation}
h^I(\tau)=\overline{p}^I+p^I\tau,\qquad
h_I(\tau)=\overline{q}_I+q_I\tau.
\end{equation}
In the horizon-limit we have 
\begin{equation}
\lim_{\tau\to\infty} \tilde{x}^a =\frac{p^I q_I - 2 p^aq_a}{2(p^bp^c-p^0q_a)},\qquad
\lim_{\tau\to\infty} \tilde{y}^a =\frac{\sqrt{I_4}}{2(p^bp^c-p^0q_a)}.
\end{equation}
In order to obtain $\vert\tilde{\Psi}(\tau)\rangle$ on the horizon,
we have to apply $(P\otimes P\otimes P)(S_3\otimes S_2\otimes S_1)$ on the charge vector 
and taking the limit $\tau\to\infty$.
For notational simplicity we introduce the abbreviation
\begin{equation}
\vert\Psi\rangle\equiv\lim_{\tau\to\infty}\vert\Psi(\tau)\rangle.
\end{equation}
Then using the identity
\begin{equation}
\label{hasznossag}
4(p^2p^3-p^0q_1)(p^3p^1-p^0q_2)(p^1p^2-p^0q_3)=
(p^0)^2I_4 + (2p^1p^2p^3 - p^0p^I q_I)^2
\end{equation}
we get
\begin{equation}
\vert\tilde{\Psi}\rangle = \frac{-i}{2} \frac{I_4^{\frac 1 4}}{\sqrt{\beta^2 + \alpha^2}}
\Bigl[\alpha(\vert000\rangle+\vert110\rangle+\vert101\rangle+\vert011\rangle)
+i\beta(\vert111\rangle+\vert001\rangle+\vert010\rangle+\vert100\rangle)  \Bigr].
\end{equation}
The discrete Fourier transform of this state is
\begin{equation}
\vert\Psi\rangle =I_4^{\frac 1 4}\frac{1}{\sqrt{2}}
\left[\frac{\beta -i\alpha}{\sqrt{\beta^2 + \alpha^2}} \vert000\rangle
-\frac{\beta +i\alpha}{\sqrt{\beta^2 + \alpha^2}} \vert111\rangle\right],
\label{lab}
\end{equation}
with
\begin{equation}
\label{alphabeta}
\alpha = \sqrt{|I_4|} p^0,\qquad
\beta = 2p^1p^2p^3 - p^0p^I q_I,
\end{equation}
in accordance with our previous results\cite{Levay1} and Eqs.~(\ref{distill}) and (\ref{fazis}).
It is important to realize at this point that our state at the horizon can alternatively
be written as
\begin{equation}
\vert\Psi\rangle=\vert Z\vert\frac{1}{\sqrt{2}}\Bigl[e^{-i\arg(Z)}\vert 000\rangle -
e^{i\arg(Z)}\vert 111\rangle\Bigr],
\label{centalt}
\end{equation}
where the quantities $\vert Z\vert$ and $\arg(Z)$ now refer to the magnitude and phase of the central charge at the horizon.
Recall that $\arg(Z(\tau))$ satisfies the following equation
\begin{equation}
\frac{d}{d\tau}\arg(Z(\tau))+{\cal A}(\tau)=0,\qquad {\cal A}=\frac{1}{2}\sum_{a=1}^3\frac{dy^a}{y^a},
\label{berry}
\end{equation}
i.e.~${\cal A}$ is the K\"ahler connection.
Hence the relative phase factors that show up in the GHZ states of Eq.~(\ref{distill}) and (\ref{lab})
are just the attractor values for the phase of the central charge governed by Eq.~(\ref{berry}).

\subsection{The non-BPS $Z=0$ case}

The non-BPS $Z=0$ solutions\cite{Scherbakov,stu} can be obtained from  the $\frac{1}{2}$-BPS ones
simply by changing the sign of any two imaginary parts of the moduli.
This yields the change of the (\ref{bpsfeltetelek}) $\frac{1}{2}$-BPS constraints
\begin{equation}
I_4>0,\qquad
p^ap^b-p^0q_c>0,\qquad
p^bp^c-p^0q_a<0,\qquad
p^cp^a-p^0q_b<0.
\label{nbpsZnullfeltetelek}
\end{equation}
During the calculation of $\vert \tilde{\Psi}\rangle$ the moduli appear only in the $S_a$ matrices.
We can carry out the sign flip of some $\tilde{y}^a$ with the $\sigma_3$ Pauli matrix:
\begin{equation}
-\sigma_3 S_a=
\frac{1}{\sqrt{\tilde{y}^a}}\begin{pmatrix}
-\tilde{y}^a&0\\
-\tilde{x}^a&1
\end{pmatrix}.
\end{equation}
Because of this $\vert\tilde{\Psi}_{23}\rangle=(\sigma_3\otimes\sigma_3\otimes I)  \vert\tilde{\Psi}\rangle$
and $\vert\Psi_{23}\rangle=(H\otimes H\otimes H)\vert\tilde{\Psi}_{23}\rangle
=(\sigma_1\otimes\sigma_1\otimes I)  \vert\Psi\rangle$,
where the indices of $\vert\Psi_{23}\rangle$ denote which moduli have been conjugated.
By virtue of these observations we get
\begin{align}
\vert\tilde{\Psi}_{12}\rangle &= \frac{-iI_4^{\frac 1 4}}{2\sqrt{\beta^2 + \alpha^2}}
\Bigl[\alpha(\vert000\rangle-\vert110\rangle-\vert101\rangle+\vert011\rangle)
        +i\beta(\vert111\rangle-\vert001\rangle-\vert010\rangle+\vert100\rangle)  \Bigr],\\
\vert\tilde{\Psi}_{23}\rangle &= \frac{-iI_4^{\frac 1 4}}{2\sqrt{\beta^2 + \alpha^2}}
\Bigl[\alpha(\vert000\rangle+\vert110\rangle-\vert101\rangle-\vert011\rangle)
        +i\beta(\vert111\rangle+\vert001\rangle-\vert010\rangle-\vert100\rangle)  \Bigr],\\
\vert\tilde{\Psi}_{13}\rangle &=\frac{-iI_4^{\frac 1 4}}{2\sqrt{\beta^2 + \alpha^2}}
\Bigl[\alpha(\vert000\rangle-\vert110\rangle+\vert101\rangle-\vert011\rangle)
        +i\beta(\vert111\rangle-\vert001\rangle+\vert010\rangle-\vert100\rangle)  \Bigr].
\end{align}
The discrete Fourier transform of these states is
\begin{align}
\vert\Psi_{12}\rangle &= I_4^{\frac 1 4}\frac{1}{\sqrt{2}}
\left[\frac{\beta -i\alpha}{\sqrt{\beta^2 + \alpha^2}} \vert011\rangle
     -\frac{\beta +i\alpha}{\sqrt{\beta^2 + \alpha^2}} \vert100\rangle\right],\\
\vert\Psi_{23}\rangle &= I_4^{\frac 1 4}\frac{1}{\sqrt{2}}
\left[\frac{\beta -i\alpha}{\sqrt{\beta^2 + \alpha^2}} \vert110\rangle
     -\frac{\beta +i\alpha}{\sqrt{\beta^2 + \alpha^2}} \vert001\rangle\right],\\
\vert\Psi_{13}\rangle &= I_4^{\frac 1 4}\frac{1}{\sqrt{2}} 
\left[\frac{\beta -i\alpha}{\sqrt{\beta^2 + \alpha^2}} \vert101\rangle
     -\frac{\beta +i\alpha}{\sqrt{\beta^2 + \alpha^2}} \vert010\rangle\right].
\end{align}
Note that $\alpha$ and $\beta$ are the same as in the $\frac 1 2$-BPS case however, now the charge configuration should be compatible with the restrictions of Eq.~(\ref{nbpsZnullfeltetelek}).

We can also write these states in the form of Eq.~(\ref{centalt}).
For example singleing out the first qubit we get 
\begin{equation}
 \vert\Psi_{23}\rangle=\vert Z_1\vert\frac{1}{\sqrt{2}}\Bigl[e^{i\arg(Z_1)}\vert 110\rangle -
  e^{-i\arg(Z_1)}\vert 001\rangle\Bigr],
\label{centaltspol}
\end{equation}
where
\begin{equation}
Z_a\equiv D_{\hat{a}}Z=-2iy^aD_a Z,
\end{equation}
\noindent
for the definition of $D_aZ$ see Eq.~(\ref{kovder}).
Here as in Eq.~(\ref{centalt}) the quantities $\vert Z_1\vert$ and $\arg(Z_1)$ refer to their attractor values.
Generally these quantities are $\tau$ dependent. For example $\arg(Z_1(\tau))$ satisfies the following equation
\begin{equation}
\frac{d}{d\tau}\arg(Z_1(\tau))+{\cal A}_1(\tau)=0,\qquad {\cal A}_1=\frac{1}{2}
\left(\frac{dy^1}{y^1}
-\frac{dy^2}{y^2}-\frac{dy^3}{y^3}\right).
\label{berry1}
\end{equation}
Hence 
just like the phase of the central charge $\arg(Z)$ for $\frac{1}{2}$-BPS solutions
for this non-BPS case the phase $-\arg(Z_1(\tau))$ flows to a value
$\arctan(\alpha/\beta)$
as determined by the expressions in Eq.~(\ref{alphabeta}).
The important difference here is the fact that in this case we have a different charge configuration which should now respect the non-BPS constraints of Eq.~(\ref{nbpsZnullfeltetelek}).
Clearly, similar results hold for the second and third qubits playing a special role.
Notice also that the quantities $\vert Z_a\vert$ where $a=1,2,3$ occurring in the expressions of  
the three-qubit states like Eqs.~(\ref{centalt}) and (\ref{centaltspol})
are just the attractor values of the {\it fake superpotential}
\begin{equation}
{\cal W}_a(\tau)=\vert Z_a(\tau)\vert.
\end{equation}
For the $\frac{1}{2}$-BPS case a similar role is played by the quantity ${\cal W}(\tau)\equiv\vert Z(\tau)\vert$.
Since
\begin{equation}
Z(\tau)=-\sqrt{2}{\Psi}_{111},\qquad Z_1=\sqrt{2}{\Psi}_{110},\qquad
Z_2(\tau)=\sqrt{2}{\Psi}_{101},\qquad Z_3=\sqrt{2}{\Psi}_{011}
\end{equation}
with the remaining amplitudes arising by complex conjugation (see Eq.~(\ref{conj}))
we see that the fake superpotential in the relevant cases is related to the magnitudes of those amplitudes which are not dying out as the corresponding BPS or non-BPS flow approaches the horizon.

\subsection{The non-BPS $Z\neq 0$ case}

The general non-BPS $Z\neq0$ case\cite{stu} is extremely different.
The general attractor flow solution is
\begin{align}
\label{nB1}
\exp(-4U(\tau)) &= h_0(\tau)h_1(\tau)h_2(\tau)h_3(\tau)-b^2,\nonumber\\
\tilde{x}^a(\tau) &=\frac{\varsigma_a\nu_a^2C^a_1+(\varsigma_a-\varrho_a)\nu_aC^a_2-\varrho_aC^a_3}
{\nu_a^2C^a_1+2\nu_aC^a_2 +C^a_3},\\
\tilde{y}^a(\tau) &=\frac{(\varsigma_a+\varrho_a)2\nu_aC_4 }
{\nu_a^2C^a_1+2\nu_aC^a_2 +C^a_3},\nonumber
\end{align}
where 
\begin{align}
\nu_a &= \nu\ee^{\alpha_a},\qquad
\nu = \left(\frac{\beta + \alpha}{\beta - \alpha}\right)^{\frac{1}{3}},\\
\label{vsig}
\varsigma_a &=\frac{\sqrt{-I_4}+(p^I q_I-2p^aq_a) }{2(p^bp^c-p^0q_a)},\\
\label{vrho}
\varrho_a   &=\frac{\sqrt{-I_4}-(p^I q_I-2p^aq_a) }{2(p^bp^c-p^0q_a)},
\end{align}
are charge-dependent constants with 
$\alpha$ and $\beta$ given by Eq.~(\ref{alphabeta}).
The $\alpha_a$ real constants satisfying the constraint
\begin{equation}
\alpha_1+\alpha_2+\alpha_3=0,
\end{equation}
\noindent
account for the flat directions\cite{flat,stu}.
The harmonic functions  now defined as
\begin{equation}
h_I (\tau)= b_I+(-I_4)^{\frac{1}{4}}\tau,
\end{equation}
giving rise to the quantities
\begin{align}
\label{nX1}
C^a_1&=h_bh_c+h_0h_a+2b,\\
C^a_2&=h_bh_c-h_0h_a,\\
C^a_3&=h_bh_c+h_0h_a-2b,\\
\label{nX4}
C_4&=\exp(-2U) = \sqrt{h_0h_1h_2h_3-b^2},
\end{align}
also making their presence in Eqs.~(\ref{nB1}).

One can obtain the non-BPS seed solution\cite{seed} 
investigated in Sec.~\ref{secSeedFlow} and \ref{secSeedDist}
as a special case of the general non-BPS $Z\neq0$ solution with the parameters 
$\varsigma_a=\varrho_a=\sqrt{\frac{-q_0p^a}{p^bp^c}}$,
$\nu=1=\nu_a$, 
$\alpha_a=0$.
Here $b=B/G_N$ is the undressed version of the $B$ field of the seed sollution.

Now we can present the horizon-limit of $\Psi$,
with some calculational steps needed for its derivation are left for the Appendix
\begin{multline}
\label{tPsinBH1}
\vert\tilde\Psi\rangle=
\frac{1}{2}
\frac{ (-I_4)^{\frac1 4} }
{\sqrt{ (\beta^2-\alpha^2)
\sgn(\nu)\cosh(\alpha_3+\varphi)\cosh(\alpha_2+\varphi)\cosh(\alpha_1+\varphi)}}\cdot\\
\cdot\Bigl[ 
 \eta_0\vert000\rangle
+\eta_1\vert110\rangle
+\eta_2\vert101\rangle
+\eta_3\vert011\rangle
+\kappa_0\vert111\rangle
+\kappa_1\vert001\rangle
+\kappa_2\vert010\rangle
+\kappa_3\vert100\rangle\Bigr],
\end{multline}
where
\begin{align}
\label{tPsinBH2}
\eta_0 &= -i\alpha,&\qquad
\eta_a &= i\left(-\beta\sinh(-\alpha_a+2\varphi) + \alpha \cosh(-\alpha_a+2\varphi)\right),\\
\kappa_0 &= -\beta,&\qquad
\kappa_a &= \sgn(\nu)\left(\beta\cosh(\alpha_a+\varphi) - \alpha \sinh(\alpha_a+\varphi)\right),
\end{align}
and let $\nu = \sgn(\nu)\ee^{\varphi}$:
\begin{equation}
\varphi = \ln\vert\nu\vert.
\end{equation}

Let us now consider some special charge-configurations giving rise to non-BPS $Z\neq0$ solutions.
For the $D0-D4$ configuration only the charges $q_0$ and $p^a$ are switched on.
We consider the case when $q_0<0$ and $p^a>0$.
Then $I_4=4q_0p^1p^2p^3<0$, $\alpha= 0$, $\beta=2p^1p^2p^3>0$, $\nu=1$,
and
\begin{equation}
\begin{split}
\vert\tilde\Psi\rangle=
&\frac{ (-4q_0p^1p^2p^3)^{\frac1 4} }
{2\sqrt{ \cosh(\alpha_3)\cosh(\alpha_2)\cosh(\alpha_1)}}\cdot\\
\cdot\Bigl[ \qquad\qquad
i&\sinh(\alpha_1)\vert110\rangle
+i\sinh(\alpha_2)\vert101\rangle
+i\sinh(\alpha_3)\vert011\rangle \\
-\vert111\rangle
+&\cosh(\alpha_1)\vert001\rangle
+\cosh(\alpha_2)\vert010\rangle
+\cosh(\alpha_3)\vert100\rangle
\quad\Bigr].
\end{split}
\end{equation}
As a special case when $\alpha_1=\alpha_2=\alpha_3=0$ one gets
\begin{equation}
\vert\tilde\Psi\rangle=
(-4q_0p^1p^2p^3)^{\frac1 4}
\frac{1}{2}\Bigl[ 
\vert001\rangle
+\vert010\rangle
+\vert100\rangle
-\vert111\rangle
\Bigr],
\label{br1}
\end{equation}
\noindent
i.e.~we get back to the state at the horizon obtained for the seed solution
in Eq.~(\ref{attractor}).

Now we consider the dual case of the $D2-D6$ charge-configuration\cite{Soroush,Cai,stu}.
After choosing $p^0>0$, $q_a>0$, $I_4=-4p^0q_1q_2q_3<0$, $\alpha=\sqrt{\vert I_4\vert}p^0>0$, $\beta=0$, $\nu=-1$, one obtains
\begin{equation}
\begin{split}
\vert\tilde\Psi\rangle=
&\frac{ (4p^0q_1q_2q_3)^{\frac1 4} }
{2\sqrt{ \cosh(\alpha_3)\cosh(\alpha_2)\cosh(\alpha_1)}}\cdot\\
\cdot\Bigl[
-i\vert000\rangle
+i&\cosh(\alpha_1)\vert110\rangle
+i\cosh(\alpha_2)\vert101\rangle
+i\cosh(\alpha_3)\vert011\rangle\\
+&\sinh(\alpha_1)\vert001\rangle
+\sinh(\alpha_2)\vert010\rangle
+\sinh(\alpha_3)\vert100\rangle
\qquad\Bigr].
\end{split}
\label{idesuss}
\end{equation}
Specially, when $\alpha_1=\alpha_2=\alpha_3=0$
\begin{equation}
\vert\tilde\Psi\rangle=
i
(4p^0q_1q_2q_3)^{\frac1 4}
\frac{1}{2}\Bigl[ 
\vert110\rangle
+\vert101\rangle
+\vert011\rangle
-\vert000\rangle
\Bigr].
\label{br2}
\end{equation}
Comparing Eqs.~(\ref{br1}) and (\ref{br2}) one can see that for vanishing flat directions the $D0-D4$ amplitudes are {\it real} and the $D2-D6$ ones are purely {\it imaginary}. Moreover, these cases are dual in  the sense that they are related by the bit flip operation $\sigma_1\otimes \sigma_1\otimes \sigma_1$.
Notice also that the norms of these states give $\sqrt{-4q_0p^1p^2p^3}$ and 
$\sqrt{4p^0q_1q_2q_3}$ apart from the fact whether the flat directions are vanishing or not. These quantities multiplied by $\pi$ give the macroscopic black hole entropy\cite{Levay3}.

It is interesting to analyse the effect of the asymptotic data on such states on the horizon. More precisely we are interested in those changes that leave the entropy (i.e.~the norm of the state) invariant.
As an example let us consider the $D2-D6$ case.
In the hope to have an effect merely on the relative phases of the state at the horizon we can adjust the asymptotic values for the charges and the $\alpha_a$ parameters (flat directions) coming from the moduli.
Other information coming from the asymptotic moduli are swallowed by the attractor mechanism.

Let us change the {\it signs} of the charges $q_1, q_2$ and $q_3$, in such a way that the constraint $p^0q_1q_2q_3>0$ is not changed.
Then one can show\cite{Levay3} that the possibilities for $\vert\tilde{\Psi}\rangle$ are 
\begin{equation}
\vert\tilde\Psi\rangle_{m_3m_2m_1}=
i
(4p^0q_1q_2q_3)^{\frac1 4}
\frac{1}{2}\Bigl[
m_1\vert110\rangle
+m_2\vert101\rangle
+m_3\vert011\rangle
-\vert000\rangle
\Bigr].
\label{br3}
\end{equation}
where
\begin{equation}
(m_3,m_2,m_1)\in\{(+++), (+--),(-+-),(--+)\}.
\end{equation}
Hence although these changes are not affecting the black hole entropy,
they have an effect on the particular form of the state. As one can check the possible changes giving rise to the four states  
of Eq.~(\ref{br3}) can be represented by
phase flip error 
operators as
\beq
\sigma_3\otimes \sigma_3\otimes I,\qquad
\sigma_3\otimes I\otimes \sigma_3,\qquad
I\otimes\sigma_3\otimes \sigma_3,\qquad
\eeq
\noindent
where for example
\beq 
\vert\tilde{\Psi}\rangle_{--+}=(\sigma_3\otimes\sigma_3\otimes I)
\vert\tilde{\Psi}\rangle_{+++}.
\eeq
\noindent
Alternatively we can apply the corresponding bit flip error operators containing $\sigma_1$ on the Fourier transformed 
states. 
Notice that these four states are all invariant under $\sigma_3\otimes \sigma_3\otimes \sigma_3$. The result of this is that the subspace spanned by these states is invariant under an {\it arbitrary number} of phase flips (or bit flips for the Fourier transformed subspace.) We can thus conclude that the effect of the change of this particular type of asymptotic data on the state at the horizon is the appearance of phase or bit flip {\it errors} associated with an invariant subspace spanned by the states of Eq.~(\ref{br3}).

Let us now remain in the non-BPS $Z\neq 0$ charge orbit and fix the signs of the $D2-D6$ charges, but this time let us change the asymptotic values for the parameters of the flat directions from $\alpha_a\equiv 0$ to 
$\alpha_a\neq0$, $\alpha_1+\alpha_2+\alpha_3=0$.
In this case we see that the uniform structure of Eq.~(\ref{br2}) deteriorates via the schematic transformation rule (neglecting the normalization factor)
\begin{equation}
\vert 000\rangle\mapsto \vert 000\rangle,\qquad\vert 110\rangle\mapsto \cosh{\alpha_1}\vert 110\rangle-i\sinh{\alpha_1}\vert 001\rangle,\qquad \text{e.t.c.}.
\end{equation}
Let us denote this new state i.e.~the one of Eq.~(\ref{idesuss}) 
by $\vert\tilde{\Psi}\rangle_{\alpha_3\alpha_2\alpha_1}$.
Then one can show that
\beq
\vert\tilde{\Psi}\rangle_{\alpha_3\alpha_2\alpha_1}=(E_3\otimes E_2\otimes E_1)\vert\tilde{\Psi}\rangle_{+++}
\label{ujerror}
\eeq
\noindent
where
\beq
E_a\equiv\frac{1}{\sqrt{\cosh\alpha_a}}\begin{pmatrix}1&0\\i\sinh\alpha_a&\cosh\alpha_a\end{pmatrix},\qquad \alpha_1+\alpha_2+\alpha_3=0.
\label{errorka}
\eeq
\noindent
Hence the changes on the state $\vert\tilde{\Psi}\rangle_{+++}$ originating from the flat directions 
have the obvious interpretation of errors of more general kind depending on continuously changing parameters.
Notice that we would have obtained the same state after changing the sign of the term $\cosh\alpha_a$ in the lower right corner of the matrix in Eq.~(\ref{errorka}). 
Such matrices $E_a^{\pm}\in \mathrm{GL}(2,\mathbb C)$ in the limit $\alpha_a\to 0$ result in the phase flip error operators $\sigma_3$ acting on the corresponding qubit we have already discussed.
In quantum information theory the $\mathrm{GL}(2,\mathbb C)$ operators acting on the qubits are called transformations associated with  stochastic local operations and classical communication (SLOCC)\cite{dur}.
It is amusing to see that though the error operators $E_a^{\pm}$ act locally but the constraint $\alpha_1+\alpha_2+\alpha_3=0$ (in the type IIA duality frame coming from deformations preserving the overall volume of $T^6$) refers to the fact that they are not independent. In quantum information theory such constraints usually refer to an agreement between the parties effected via the use of classical channels.
Finally in closing this subsection we note that the normalized part of the attractor state of Eq.~(\ref{br2}) i.e. $\vert\tilde{\Psi}\rangle_{+++}$ is just the one which can be used to establish a very striking version of Bell's Theorem.\cite{Mermin}

\subsection{The $D0-D6$ case}

As another special subcase of the non-BPS $Z\neq0$ one, finally we consider the $D0-D6$ solution. 
This charge-configuration can only appear in the non-BPS regime, 
because $I_4 = -(p^0q_0)^2 < 0$ independent of the signs of the charges. 
Originally, the general non-BPS $Z\neq 0$ solution was  produced by an 
$\mathrm{SL}(2,\mathbb{R})^{\otimes 3}$ U-duality transformation of the $D0-D6$ one\cite{seed,stu}. 
Due to the $\varsigma_a$, $\varrho_a$ parametrisation of this transformation, 
(see Eqs.~(\ref{vsig}) and (\ref{vrho})) 
we can not produce neither the identity transformation 
nor the transformations that bring us back to the $D0-D6$ case with different charges. 
Hence we can not simply write the $D0-D6$ charges into the corresponding formulae for the horizon-limit of $\Psi(\tau)$. 
However, 
for the calculation of $\Psi(\tau)$ on the horizon
we can directly use the original $D0-D6$ solutions\cite{stu} instead. 
\begin{align} 
\exp(-4U(\tau)) &= h_0(\tau)h_1(\tau)h_2(\tau)h_3(\tau)-b^2,\\
\tilde{x}^a(\tau)&=\frac{\nu'_aC^a_2}{C^a_3},\\ 
\tilde{y}^a(\tau)&=\frac{2\nu'_aC_4}{C^a_3}, 
\end{align} 
with the notation introduced in Eqs.~(\ref{nX1})-(\ref{nX4}), and 
\begin{equation}
\nu'_a =\left(\frac{q_0}{p^0}\right)^{\frac1 3}\ee^{\alpha_a}.
\end{equation}
Note, that on the horizon $x^a\to0$ and $y^a\to\nu'_a$.
With this moduli the calculation of $\Psi$ is much easier,
than in the general case. Finally a straightforvard calculation yields the result
\begin{equation}
\vert\tilde\Psi\rangle=
\frac{1}{\sqrt2}
(-I_4)^{\frac1 4}
\Bigl[ 
-i\vert000\rangle
+\vert111\rangle
\Bigr].
\end{equation}
Note that this state is independent of the $\alpha_a$ parameters of flat directions.

\section{Conclusions}

In this paper we have shown that in the special case of the STU model the attractor mechanism for extremal, static and spherically symmetric
BPS and non-BPS black hole solutions can be cast in a form of a distillation procedure of entangled three-qubit states of special kind on the horizon.
Such states are belonging to the so called GHZ-class featuring maximum tripartite entanglement\cite{dur}.
In obtaining this result our main calculational tool was to organize
the charges, the moduli fields and the warp factor in a suitably defined three-qubit state (Eqs.~(\ref{Smatrix}), (\ref{3qubitallapot}) and (\ref{universalstate})). This state is just lying on the $\mathrm{GL}(2,\mathbb C)^{\times 3}$ orbit of a three-qubit "charge state" (Eq.~(\ref{cgamma})) fixing the charge orbit to which the particular solution belongs.
To conform with the well-known fact that the duality group of the STU model is not $\mathrm{GL}(2,\mathbb C)^{\times 3}$ but  $\mathrm{SL}(2,\mathbb R)^{\times 3}$ our state is also satisfying a reality condition (Eq.~(\ref{reality})).
On this three-qubit state the flat covariant derivatives are acting
as phase (sign) or bit flip errors (Eq.~(\ref{flatkovariantder})) depending on whether we express the sate in the computational basis or in its discrete Fourier transformed version (Eq.~(\ref{Hadamard})-(\ref{HadamardBase})).
The black hole potential can be expressed as the norm of our state (Eq.~(\ref{Norma})). 

For spherically symmetric solutions such states are also displaying an explicit dependence on the radial coordinate $r=\frac{1}{\tau}$ referring to the distance from the horizon.
By solving the equations of motion for the moduli fields and warp factor we end up with a flow of BPS or non BPS type depending on the charge configuration.
Using the explicit forms of these flows that has recently become available in the literature 
we can study the distillation procedure in detail.
In order to illustrate how this distillation becomes unfolded as we approach the horizon we have chosen the recently discovered non-BPS seed solution.

For such solutions we observed that the charge, moduli, and warp factor dependent state
of Eq.~(\ref{universalstate}) satisfies a system of first order differential equations (Eqs.~(\ref{E2}) and (\ref{E4})) featuring the fake superpotential (Eq.~(\ref{fakewarp})).
This observation conforms with the recent results on the first order formalism relating supergravity flows to geodesic motion on the moduli space of the $3D$ dimensionally reduced theory\cite{fake,Stelle}. 
In the light of this connection it would be nice to elaborate further on this point and establish an entanglement based understanding of some of these results. 

For the non-BPS seed solutions we managed to demonstrate how a standard GHZ state at the horizon emerges from a state characterizing the flow at the asymptotically Minkowski region.
The attractor mechanism in this picture simply amounts to the fact that three amplitudes out of the seven nonequal nonvanishing ones of our three-qubit state are dying out as we approach the horizon (see Eqs.~(\ref{Minkowski}) and (\ref{attractor})).
The remaining amplitudes have the same magnitudes related to the macroscopic black hole entropy. The relative phase factors of these amplitudes are merely signs reflecting the srtucture of the fake superpotential\cite{fake,stu}.

In this paper we also conducted a detailed study on the structure of
"attractor states". By this term we denote the particular states we obtain from our $\tau$ dependent ones after performing the $\tau\to\infty$ limit.
We have shown that the $\frac{1}{2}$-BPS and non-BPS $Z=0$ cases are
very similar. The attractor states are of canonical GHZ form with the relative phases related to the phases of the central charge $Z$ or the phases of the quantities $Z_a$, $a=1,2,3$ which are just the flat covariant derivatives of $Z$ (Eqs.~(\ref{centalt}) and (\ref{centaltspol})). 
We observed that in these cases and also in the case of the $Z\neq 0$ seed solutions the fake superpotential is related to those amplitudes of our $\tau$ dependent states which are {\it not} dying out as we approach the horizon.

The new subtlety arising in the non-BPS $Z\neq 0$ case is the appearance of flat directions.
As it is known for the $D0-D6$ case the parameters labelling the flat directions are related to deformations of the volumes of the three tori $T^2$ preserving the overall volume of $T^6$ (in the type IIA duality frame).
We have found that for this charge configuration the flat directions are not making there presence in the corresponding attractor state.
(Though they do appear in the particular form of the attractor values of the moduli.)
However, for the most general charge configuration flat directions do appear in the attractor states. 
In the special cases of the $D0-D4$ and $D2-D6$ systems we have shown that
the effect of the flat directions is to deteriorate the uniform structure of the corresponding attractor states obtained by starting the flow not in any of the flat directions.
It is known\cite{Levay3} that the effect of changing the signs of the $D2$ and $D4$ charges asymptotically results in phase or bit flip errors on the attractor states.
By virtue of this the presence of flat directions adds an additional twist to this picture.
In particular we have demonstrated that flat directions can entertain the possible interpreatation as errors of more general type (i.e.~ones depending on continuously changing parameters) acting on attractor states.

Now we comment on the possible physical relevance of our three-qubit states. Obviously our compressing of the variuos ingredients of the STU model in a three-qubit state at this stage is merely a nice way of understanding the structure of BPS and non-BPS solutions in the STU model.
Notice however, that the attractor states are always just the ones
that are connected to the structure of the fake superpotential.
Indeed, the fake superpotential in our examples turned out to be 
related to those amplitudes of our three-qubit states which are {\it not dying out} during the process of distillation. 
It is known that upon quantization of the radial evolution of the moduli\cite{Pioline} results in a semiclassical wave function the phase of which is featuring the quantity $e^U{\cal W}$ which is just formed out of the aforementioned amplitudes of our three-qubit state of Eq.~(\ref{universalstate}). (See also Eqs.~(\ref{fakewarp}), (\ref{fake}) and (\ref{explfake}).)
This observation might give a clue towards getting a deeper insight into the physical meaning of our GHZ-like states.

Finally notice that although being very special, the STU model captures the essential features 
also of extremal black holes in the $N=4,8$ theories. Moreover many of its features generalize well to other black hole solutions (such as those arising from CY compactifications).
In this respect we just remark that the maximal $N=8$, $d=4$ supergravity has {\it seven} STU subsectors corresponding to its consistent truncations. In the corresponding extremal black hole solution context this observation has already been related to systems exhibiting tripartite entanglement of seven qubits\cite{DF1,Levay2}.
It would be interesting to study distillation issues for this more general scenario using the ideas as developed in this paper.

\appendix

\section{Calculating $\Psi(\tau)$ on the horizon (non-BPS $Z\neq0$ case)}
\label{nbpsapp}

In this appendix we outline the main steps leading to the
explicit expression of $\Psi$ on the horizon.
First recall the explicit form of the non-BPS $Z\neq0$ attractor flow of Eqs.~(\ref{nB1})-
(\ref{nX4}).
Using the (\ref{vsig}) and (\ref{vrho}) forms of $\varsigma_a$ and $\varrho_a$,
we can alternatively write this flow as
\begin{align}
\label{xX}
\tilde{x}^a(\tau)&= \frac{p^I q_I - 2 p^aq_a}{2(p^bp^c-p^0q_a)} + \frac{\sqrt{-I_4} }{2(p^bp^c-p^0q_a)}C^a_x(\tau),\\
\label{yY}
\tilde{y}^a(\tau)&= \frac{\sqrt{-I_4}}{2(p^bp^c-p^0q_a)} C^a_y(\tau).
\end{align}
Here the $\tau$ dependent terms are
\begin{align}
C^a_x(\tau)&=\frac{\nu_a^2C^a_1-C^a_3}{\nu_a^2C^a_1+2\nu_aC^a_2 +C^a_3},\\
C^a_y(\tau)&=\frac{4\nu_aC_4}{\nu_a^2C^a_1+2\nu_aC^a_2 +C^a_3}.
\end{align}
Since the moduli can be written as $\tilde{x}^a(\tau)=A^a + B^a C^a_x(\tau)$ and  $\tilde{y}^a(\tau)=B^a C^a_y(\tau)$,
the transformation $S_a$ of Eq.~(\ref{Smatrix}) can be expressed as
\begin{equation}
S_a=\frac{1}{\sqrt{y^a}}\mathcal{C}_a\mathcal{B}_a\mathcal{A}_a
\end{equation}
where
\begin{equation}
\mathcal{C}_a=\begin{pmatrix}
C^a_y & 0\\
-C^a_x & 1
\end{pmatrix},\qquad
\mathcal{B}_a=\begin{pmatrix}
 B^a & 0\\
 0 & 1
\end{pmatrix},\qquad
\mathcal{A}_a=\begin{pmatrix}
 1 & 0\\
 -A^a & 1
\end{pmatrix}.
\end{equation}

After these preliminaries the transformation $S_3\otimes S_2\otimes S_1$ can be carried out in three steps.
First in order to use the (\ref{alphabeta}) definition of $\alpha$ and $\beta$ also
in this non-BPS case, the identity (\ref{hasznossag}) have to be changed as 
\begin{equation}
4(p^2p^3-p^0q_1)(p^3p^1-p^0q_2)(p^1p^2-p^0q_3)=\beta^2-\alpha^2
\end{equation}
due to $I_4(\Gamma)<0$.
By virtue of this we can perform the first two transformation
\begin{multline}
\label{gghz}
(\mathcal{B}_3\mathcal{A}_3\otimes
 \mathcal{B}_2\mathcal{A}_2\otimes
 \mathcal{B}_1\mathcal{A}_1)\vert\Gamma\rangle=\\
=\frac{1}{\sqrt 8}
\frac{-I_4}{ \beta^2 - \alpha^2 }
\Bigl[\alpha(\vert000\rangle+\vert011\rangle+\vert101\rangle+\vert110\rangle)
+\beta(\vert111\rangle+\vert100\rangle+\vert010\rangle+\vert001\rangle)\Bigr]=\\
=\frac{-I_4}{ \beta^2 - \alpha^2 }
\left[\frac{1}{2}(\beta+\alpha)\vert\tilde{0}\tilde{0}\tilde{0}\rangle
-\frac{1}{2}(\beta-\alpha)\vert\tilde{1}\tilde{1}\tilde{1}\rangle\right]
\end{multline}
where $\vert\tilde{0}\rangle$ and $\vert\tilde{1}\rangle$ are the (\ref{HadamardBase}) Hadamard transformed states. 
On this latter form the transformation $(\mathcal{C}_3\otimes \mathcal{C}_2\otimes \mathcal{C}_1)$
acts readily.
The $\tau$-dependency appears only in $\mathcal{C}_a$. In the horizon-limit we obtain
\begin{equation}
\lim_{\tau\to\infty} C^a_x = \frac{\nu_a^2-1}{\nu_a^2+1}
=\frac{\frac{1}{2}\left(\nu_a-\frac{1}{\nu_a}\right)}{\frac{1}{2}\left(\nu_a+\frac{1}{\nu_a}\right)},\qquad
\lim_{\tau\to\infty} C^a_y = \frac{2\nu_a}{\nu_a^2+1}
=\frac{1}{\frac{1}{2}\left(\nu_a+\frac{1}{\nu_a}\right)},
\end{equation}
with these formulae we get the result of (\ref{tPsinBH1})-(\ref{tPsinBH2}).


\end{document}